\RequirePackage{fix-cm}
\documentclass{svjour3}                     
\smartqed  
\usepackage{graphicx}
\usepackage[top=30truemm,bottom=30truemm,left=25truemm,right=25truemm]{geometry}
\usepackage{color}
\usepackage{amsmath}
\usepackage{lscape}
\usepackage{amsfonts}
\usepackage{bm}

\begin{document}

\title{{Assessment of supervised machine learning methods for fluid flows}}

\titlerunning{{Assessment of supervised machine learning methods for fluid flows}}

\author{Kai Fukami \and Koji Fukagata \and  Kunihiko Taira
}

\institute{Kai Fukami, Koji Fukagata \at
              Department of Mechanical Engineering, Keio University, Yokohama 223-8522, Japan
              \\
              \email{kai.fukami@keio.jp}           
           \and
           Kunihiko Taira  \at
              Department of Mechanical and Aerospace Engineering, University of California, Los Angeles, CA 90095, USA
}

\date{Received: \today / Accepted: }

\maketitle

\begin{abstract}
We apply supervised machine learning techniques to a number of regression problems in fluid dynamics.  
Four machine learning architectures are examined in terms of their characteristics, accuracy, computational cost, and robustness for canonical flow problems.  
We consider the estimation of force coefficients and wakes from a limited number of sensors on the surface for flows over a cylinder and NACA0012 airfoil with a Gurney flap.  
The influence of the temporal density of the training data is also examined.  
Furthermore, we consider the use of convolutional neural network in the context of super-resolution analysis of two-dimensional cylinder wake, two-dimensional decaying isotropic turbulence, and three-dimensional turbulent channel flow.  
In the concluding remarks, we summarize on findings from a range of regression type problems considered herein.
\keywords{Supervised Machine Learning, Wake Dynamics, Turbulence}
\end{abstract}

\section{Introduction}

In recent years, machine learning approaches have been successfully applied to a range of nonlinear and chaotic fluid flow problems.  
These applications have leveraged the powerful capability of supervised machine learning to serve as an universal approximator for the nonlinear relationship between input and output data \cite{Kutz2017,BNK2019,BEF2019,Kreinovich1991,Hornik1991,Cybenko1989,BFK2018}.  
Within a short time frame, these efforts have produced a vast collection of studies with a wide selection of machine learning techniques.  
For an aspiring fluid mechanician, a naive dive into the ocean of machine learning models may be not only a bit intimidating but also nauseating.  
The objective of this paper is to present some of the notable techniques and provide guidelines on their uses for canonical regression problems for laminar and turbulent flows. 

Machine learning has been applied to problems, including turbulence modeling for Reynolds Averaged Navier--Stokes \cite{LKT2016,DIX2019} and large-eddy simulations \cite{GH2017,MS2017}, modeling the dynamics of fluid flows \cite{SM2018,LW2019,SGASV2019,MFF2019}, flow estimation \cite{SP2019,LMB2018} and reconstruction \cite{FFT2019,FNKF2019,DCLK2019,EMYBMK2019}, and optimization of flow control designs \cite{DBN2017,RKJRC2019}.  
These studies have shown that machine learning techniques can accurately model the complex input-output relationship within nonlinear fluid flows.  
In essence, these studies benefit from the ability of machine learning techniques to extract the nonlinear input-output mapping from data.  
Although great strides have been made to date, the field of fluid mechanics has not seen concrete presentations of benchmarks and guidelines to construct machine learning models. 
As such, we can benefit from a systematic procedure to develop machine learning models appropriate for fluid dynamics. 
In the present paper, we discuss how one may select a suitable approach from a large family of machine learning models developed in the computer science field. 

\begin{figure}
	\vspace{0mm}
	\begin{center}
		\includegraphics[width=1.00\textwidth]{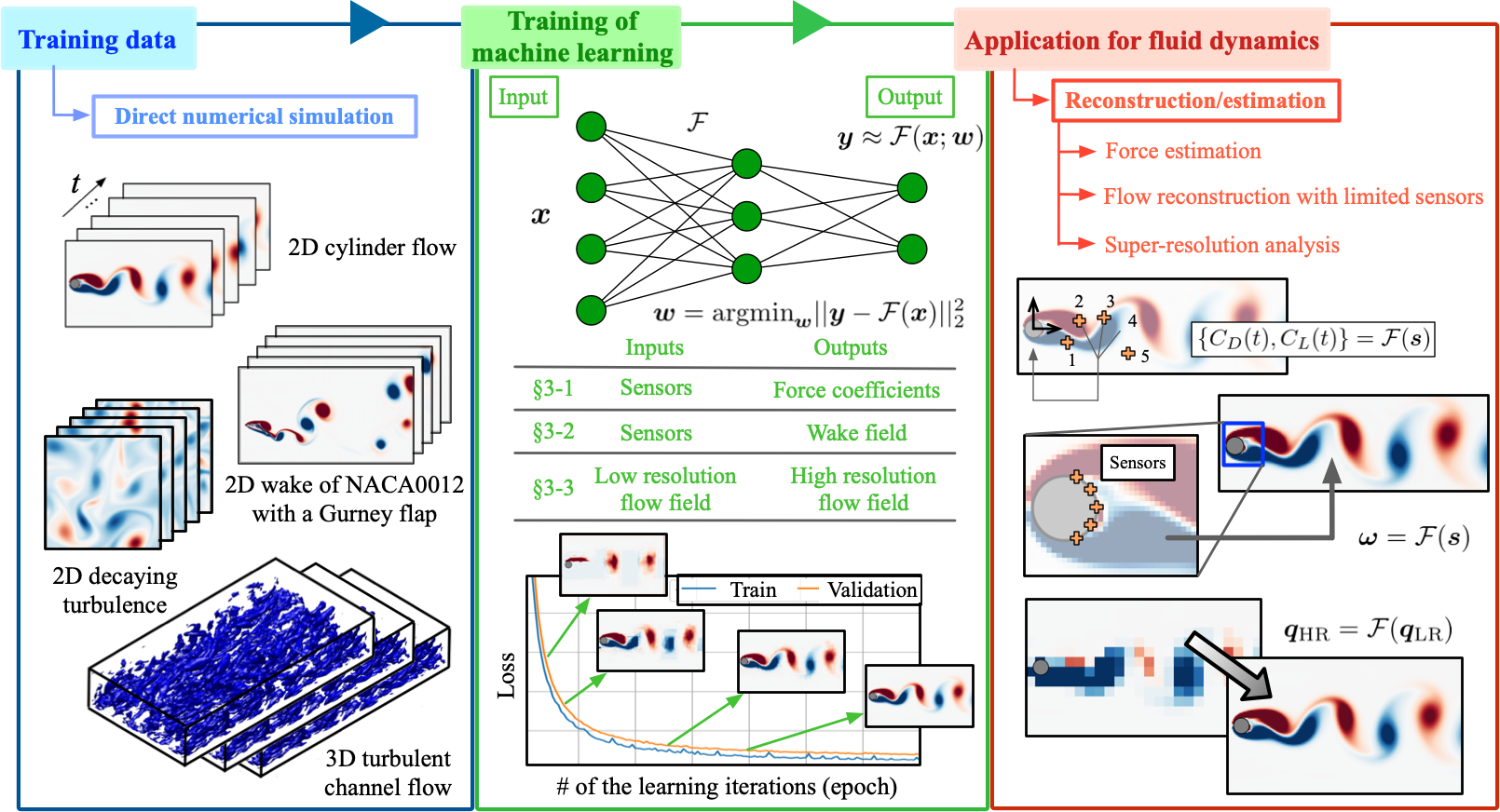}
		\caption{Overview of supervised machine learning problems for fluid flows discussed in this paper.  We focus on reconstruction and estimation tasks for direct numerical simulation data.}
		\label{fig00}
	\end{center}
\end{figure}

We consider a systematic procedure to develop supervised machine learning models and assess them on canonical fluid flow problems.   
With supervised machine learning, training data comprised of input {$\bm x$} and output (solution) {$\bm y$} data are needed.  
The machine learning models are then trained using the given data sets to seek a nonlinear mapping ${\bm y}\approx{\cal F}({\bm x};{\bm w})$.  
This can be mathematically posed as a problem to find the optimal weights $\bm w$.  
Here, we consider four machine learning architectures to assess their characteristics, accuracy, computational cost, and robustness for laminar flow problems.  
Furthermore, we also examine the use of convolutional neural network (CNN) for super-resolution analysis.  
An overview of the current study is presented in figure \ref{fig00} with illustrations of the example problems.

In section \ref{sec:MLmethods}, we introduce five machine learning models that have been considered for fluid dynamics problems: (1) multi-layer perceptron, (2) random forest, (3) support vector regression, (4) extreme learning machine, and (5) CNN.  
These models are applied to regression problems in section \ref{sec:examples}.  
Namely, these problems are (1) the estimation of drag and lift coefficients for laminar flows, (2) the estimation of laminar wake flow from limited measurements, and (3) the super-resolution analysis for coarse flow data.  
We present the procedures to develop machine learning models and guidelines to ensure the models are accurate and robust.  
At last in section \ref{sec:conclude}, we offer concluding remarks and possible extensions to currently available machine learning models.

\section{Machine learning methods}
\label{sec:MLmethods}

\begin{table}
\begin{center}
\caption{Strengths and weaknesses of the machine learning models covered in this paper.}
\def~{\hphantom{0}}
\begin{tabular}{cccc}
     \hline 
        ML models & Strengths & Weaknesses & Examples\\ \hline\hline
        MLP & Compare easily & Curse of dimensionality  & $C_D$,$C_L$ estimation (\S 3-1) \\ 
        (\S 2-1-1) & (Commonly used) &   & Wake reconstruction (\S 3-2) \\ \hline 
        RF & Interpretability & Curse of dimensionality & $C_D$,$C_L$ estimation (\S 3-1) \\
        (\S 2-1-2) & (Feature importance) &  & Wake reconstruction (\S 3-2)\\ \hline 
        SVR  & Avoids overfitting & Curse of dimensionality & $C_D$,$C_L$ estimation (\S 3-1)\\
        (\S 2-1-3) & (Max margin problem) &  & Wake reconstruction (\S 3-2)\\ \hline
        ELM  & Fast computation & Noise robustness & $C_D$,$C_L$ estimation (\S 3-1)\\
        (\S 2-1-4) &  &  &Wake reconstruction (\S 3-2) \\ \hline
        CNN & Low computational cost & Limited for uniform grid & Super resolution (\S 3-3) \\
        (\S 2-1-5) &  Handles big data &  &\\
     \hline 
    \end{tabular}
  \label{tab1}
\end{center}
\end{table}

Constructing a supervised machine learning model requires a few steps.  
First, we must prepare the training data comprised of the input $\bm x$ and output $\bm y$.
Second, a model $\cal F$ itself must be chosen based on its properties.  
Given the model and the available data set, we can then optimize the weights $\bm w$ to train the model, as illustrated in the green box of figure \ref{fig00}.  
In this paper, we consider five machine learning models $\cal F$ as presented in section \ref{sec:model}.  
We also discuss the choice of norms for the loss function needed to optimize the model weights $\bm w$ in section \ref{sec:norm} and emphasize the need for cross validation to prevent overfitting in section \ref{sec:cv}.

\subsection{Models}
\label{sec:model}

For the example problems that will be considered in this paper, we can select a machine learning model from a very large number of available models \cite{BK2019}.  
In this section, we examine five of the representative machine learning models.
Their strengths and weaknesses are summarized in table \ref{tab1}.  
We choose each of these machine learning models based on their strengths for each of the examples studied later.  
Here, we start our presentation of the models with the {\it multi-layer perceptron}, which has been widely used in fluid dynamics, yielding a large number of references.  
Next, we discuss an interpretable machine learning model, {\it random forest} that can reveal the feature importance (importance of the input influence).  
Also considered is the {\it support vector regression}, which is robust against overfitting.  
Highlighting the computational aspect, we cover the {\it extreme learning machine} founded on the shallow network and the least squares formulations.  
In this paper, these four machine learning models are compared with each other in terms of accuracy, computational cost, and robustness for canonical laminar flow problems.  
For fairness in comparison, parameters of each machine learning model are optimized as explained later. 
We also present the {\it convolutional neural network} which reveals its strength in handling high-dimensional input and output data.  Below, let us briefly discuss each aforementioned machine learning model.

\subsubsection{Multi-layer perceptron}

A multi-layer perceptron (MLP) developed by Rumelhart {\it et al}. \cite{RHW1986} is pervasive in a wide range of sciences.
The basic concept of MLP is motivated by the structure of biological neural circuits.  
An MLP has been successfully applied to diverse problems, including regression, classification, and speech recognition \cite{Domingos2012}.  
We have also seen the use of MLP for turbulence modeling, reduced order modeling, and flow reconstruction in fluid dynamics \cite{LKT2016,LW2019,MSRV2019,YH2019}.

\begin{figure}
	\vspace{0mm}
	\begin{center}
		\includegraphics[width=0.85\textwidth]{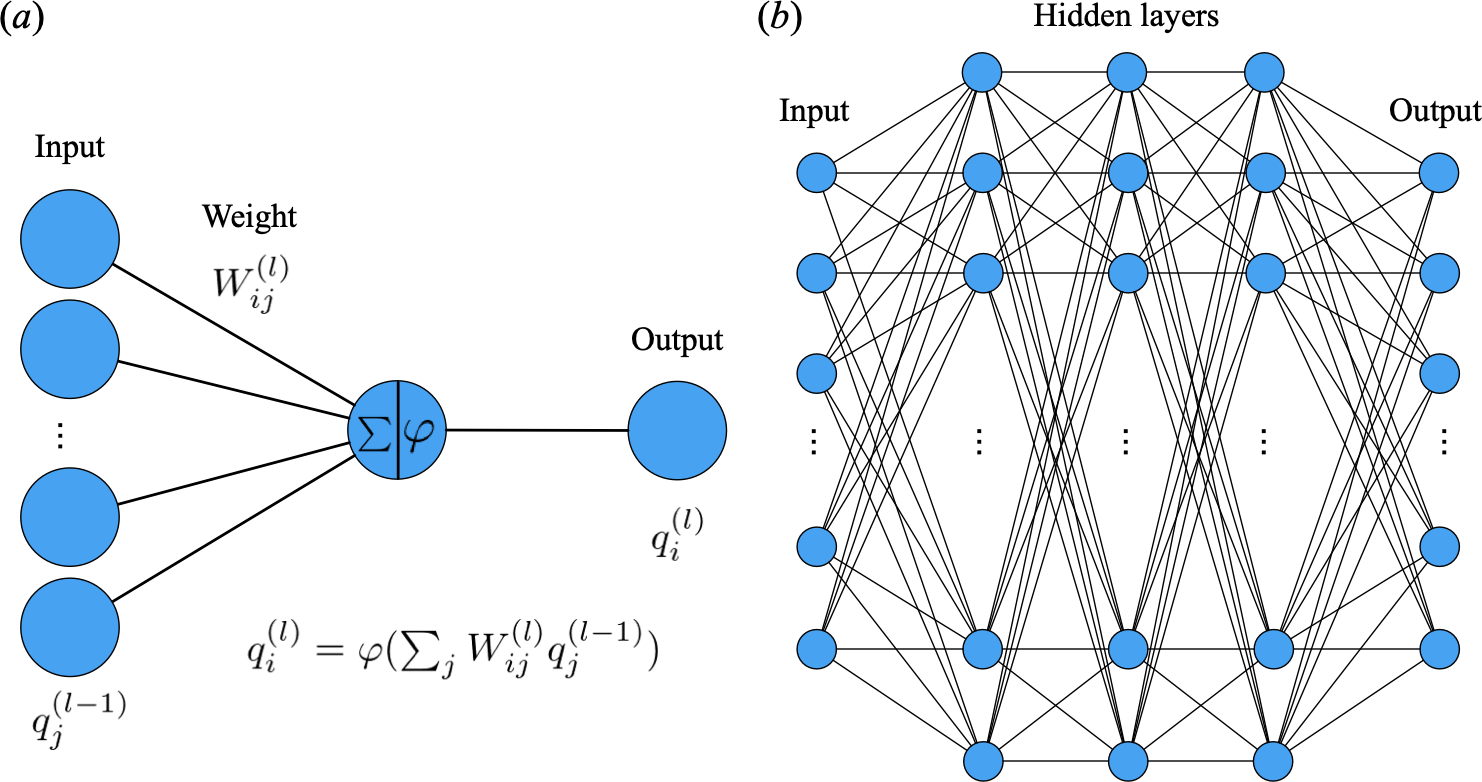}
		\caption{
		$(a)$ A perceptron. (b) Multi-layer perceptron is an aggregate of perceptrons.}
		\label{fig1}
	\end{center}
\end{figure}

\begin{figure}
	\vspace{0mm}
	\begin{center}
		\includegraphics[width=0.80\textwidth]{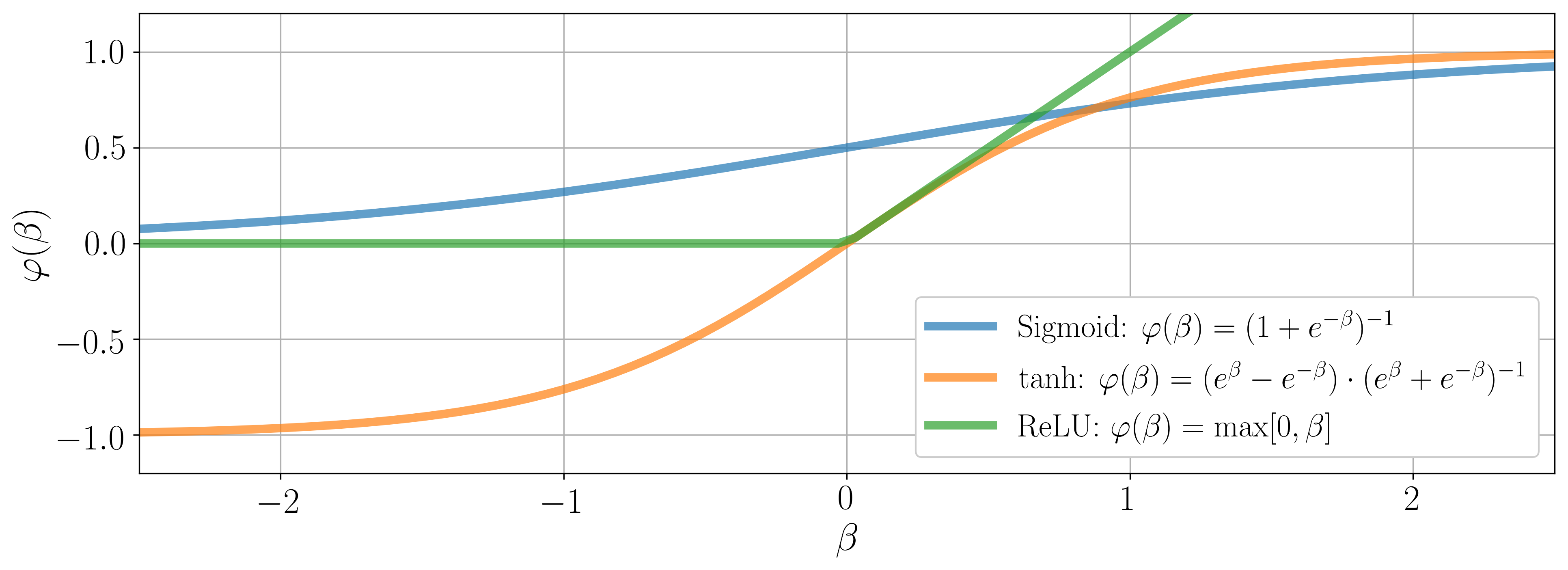}
		\caption{An example of activation functions $\varphi(\beta)$.  In the present papar, we use ReLU $\varphi(\beta)={\rm max}[0,\beta]$.}
		\label{fig3_1}
	\end{center}
\end{figure}

A minimum unit is called a perceptron as shown in figure \ref{fig1}$(a)$. 
The input data from the $(l-1){\rm th}$ layer are multiplied a weight $\bm{W}$. 
These inputs are linearly combined and passed through a nonlinear activation function $\varphi$:
\begin{eqnarray}
    q_i^{(l)}=\varphi(\sum_j W^{(l)}_{ij}q_j^{(l-1)}).
    \label{eq:1}
\end{eqnarray}
An MLP is constructed by connecting these perceptrons in layers, as illustrated in figure \ref{fig1}$(b)$.
Weights for all links $W_{ij}$ are optimized to minimize a positive-definite loss (cost) function ${E}$ by back propagation \cite{KB2014} such that ${\bm w}={\rm argmin}_{\bm w}[{E}({\bm y},{\mathcal F}({\bm x};{\bm w}))]$.

The activation function $\varphi$ is generally chosen to be monotonically increasing nonlinear functions.  
Some of the widely used functions of sigmoid $\varphi(\beta)=(1+e^{-\beta})^{-1}$, hypabolic tangent (tanh) $\varphi(\beta)=(e^{\beta}-e^{-\beta})\cdot(e^{\beta}+e^{-\beta})^{-1}$, and ReLU $\varphi(\beta)={\rm max}[0,\beta]$, are shown in figure \ref{fig3_1}.  
All of these activation functions seem to perform well for shallow networks.  
For deep networks, ReLU is known to be a good candidate \cite{NH2010}.  
Details on the specific choices and their influence on networks can be found in Haykin \cite{Haykin1998} and Nair and Hinton \cite{NH2010}.  
In the MLP formulation, we use ReLU activation function (we also use ReLU later in the extreme learning machine and convolutional neural network formulations).

Let us present the steps to optimize the MLP parameters (numbers of units and layers) for actual examples of force estimation and wake reconstruction based on wake measurements, as discussed later in sections 3.1 and 3.2.  
As mentioned above, the aim here is to find the optimal numbers of units and layers for the multi-layer perceptron.  In this case, the following procedure can be used:
\begin{enumerate}
  \item We fix the number of layers $n_{\rm layer}$ and find the number of units $n_{\rm unit}^*$ that minimizes the loss function.  Multiple cross validations should be used.
  \item We then fix the number of units $n_{\rm unit}^*$ and find the number of layers $n_{\rm layer}^*$ that minimizes the loss function.  Multiple cross validations should be used.
  \item The found parameter set $(n_{\rm unit,opt},n_{\rm layer,opt}) = (n_{\rm unit}^*,n_{\rm layer}^*)$ are used as the optimized parameters for the machine learning model.  Depending on the problem, one can iterate over steps 1 and 2.
\end{enumerate}
We also apply the same procedure to other machine learning models in this paper, as explained later.

\begin{table}
\begin{center}
\caption{{Examples of optimized parameters of the machine learning model used in this paper.}}
\def~{\hphantom{0}}
\begin{tabular}{cccc}
     \hline \hline
        ML models & 1st parameter & 2nd parameter & 3rd parameter\\ \hline
        MLP & Unit & Layer  & --- \\ \hline
        RF & Depth & \# of decision trees $N_{\rm Tree}$  & --- \\ \hline
        SVR  & Epsilon tube $e$ & Coefficient of error term $C$ & ---\\\hline
        ELM  & Unit & --- & ---\\\hline
        CNN & Layer & \# of filters & Filter size \\
     \hline \hline
    \end{tabular}
  \label{tab_rev1}
\end{center}
\end{table}

{The optimized parameters for each machine learning model of the present study are summarized in table~\ref{tab_rev1}.
The details on the choice of each parameter are discussed later.}
{Note that the proposed order here, e.g., the units first and the layer next in the case of MLP, is just an example order taken in the present study.
Users are able to change the order within the optimization process.
As the other methods for hyper parameter decision, users can also consider the use of theoretical optimization methods, e.g., hyperopt \cite{BYC2013} and Baysian optimization \cite{BCF2009,MMLMBL2019}.
}

\subsubsection{Random forest}

\begin{figure}
	\vspace{0mm}
	\begin{center}
		\includegraphics[width=0.80\textwidth]{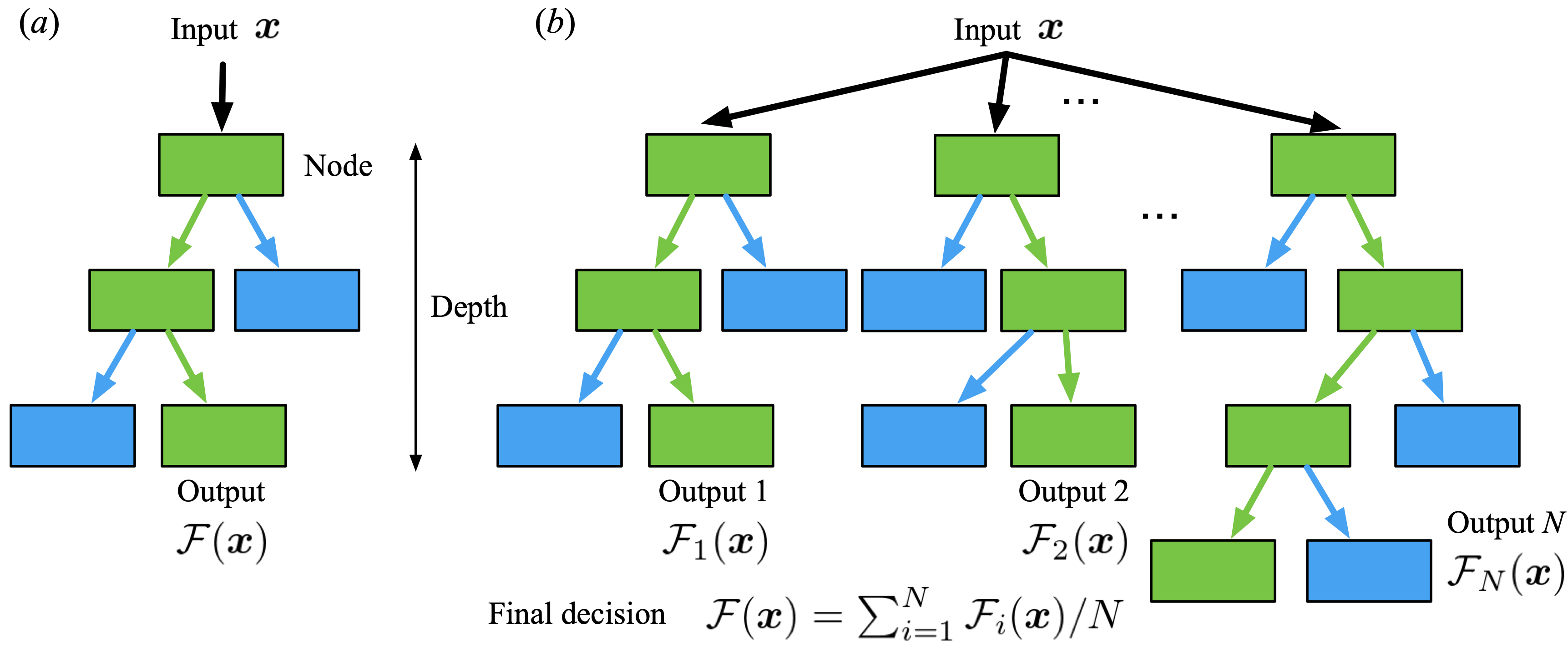}
		\caption{$(a)$ Decision tree. $(b)$ Random forest, which can be regarded as ensemble average of decision trees.}
		\label{fig2}
	\end{center}
\end{figure}

Next, let us consider a decision tree \cite{Quinlan1986} which is the basis for a random forest (RF). 
A decision tree has a flow-chart structure as shown in figure \ref{fig2}$(a)$. 
The boxes in figure \ref{fig2}$(a)$ are called nodes and represent the group of features and data.  
The learning objective of a decision tree is to seek {\it if}/{\it then}/{\it else} rules for obtaining the desired output $\bm y$, as branches (the allows in figure \ref{fig2}$(a)$).  
Let us present a simple example for understanding the construction of a decision tree: a model to estimate the amount of the water a person drinks when the temperature is $T$. 
If a person drinks a lot of water when the temperature $T$ is more than the threshold $T_{\rm th}$, the rule for water consumption would depend on $T\geq T_{\rm th}$ and $T< T_{\rm th}$.  
These rules are illustrated by the green and blue lines of figure \ref{fig2}$(a)$. 
In this example, humidity is also another factor that influences the amount of water consumption. 
We can consider an increased number of influential variables so that the accuracy of the model is improved.  
To incorporate the complexity, the depth of a decision tree can be increased as shown in figure \ref{fig2}$(a)$.
In summary, the boxes and lines of figure \ref{fig2}$(a)$ can be employed to construct rules for obtaining the output of the machine learning model ${\cal{F}}({\bm x})$.

The decision tree can be explainable because the internal structure is comprised of simple {\it if}/{\it then}/{\it else} rules.  
However, it is known that the learning process of a decision tree is unstable and prone to overfitting due to the vulnerability with respect to small variations in the data \cite{DH2007}.  
To address these issues, an ensemble learning approach \cite{OM1999} is often adopted. 
{\it Random forest} (RF) \cite{Breiman2001} is one of these ensemble learning techniques for a decision tree.  
The output value of RF is given an average score of the predictions of all decision trees, as shown in figure \ref{fig2}$(b)$.  
As the optimized parameters, we focus on the depth and number of decision trees $N_{\rm Tree}$ (vertical and horizontal direction in figure \ref{fig2}$(b)$) {as summarized in table \ref{tab_rev1}}, respectively.
{We note in passing that RF contains considerable hyper parameters, e.g., maximum depths of the tree and minimum number of samples required to be at the leaf node.
Users can also consider these parameters for implementation.}

\subsubsection{Support vector regression}

Support vector regression (SVR) \cite{SS2004} is an extension of support vector machine used for classification \cite{VL1963}.  
This method has been also utilized for various problems that are similarly tackled by MLPs and RFs \cite{Bishop2006}.
In the field of fluid dynamics, the support vector machine has been applied to turbulence modeling \cite{LT2015} and reduced-order modeling \cite{CJKMPW2019}.

\begin{figure}
	\vspace{0mm}
	\begin{center}
		\includegraphics[width=0.70\textwidth]{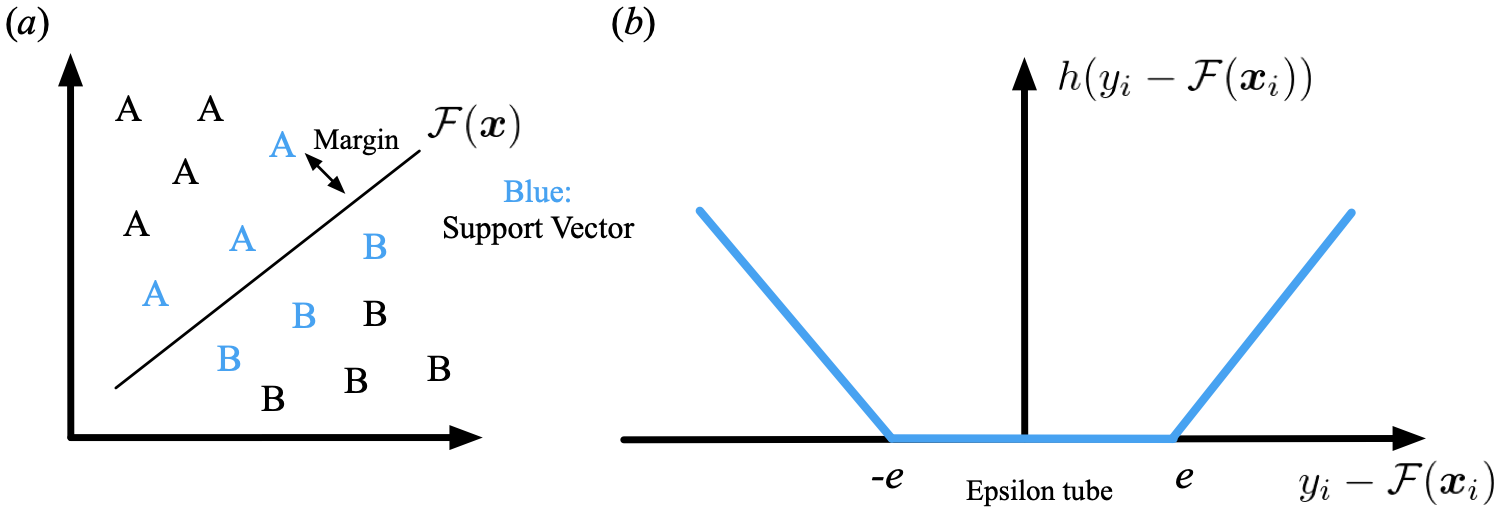}
		\caption{$(a)$ Support vector machine with two-group classification. $(b)$ The concept of epsilon tube.}
		\label{fig3}
	\end{center}
\end{figure}

Here, let us consider a two-dimensional classification problem for demonstration, as shown in figure \ref{fig3}$(a)$. 
To classify data into two groups of A and B, we utilize only the components around the borders represented as blue components called the support vector. 
Next, to avoid misclassification, the distance between the border and neighboring components should be maximized.  
This problem is called the {\it maximum margin problem}.  
We can also use the same concepts for regression problems.  
Mathematically speaking,  the objective of the learning process is to establish the function ${\cal F}({\bm x})={\bm W}^T\phi({\bm x})$ between input and output data by mapping training data into the high-dimensional feature space for regression.   
Here, the ${\bm W}$ and $\phi$ are the high-dimensional weight and map function, respectively.  
Seeking the border in figure \ref{fig3}$(a)$ corresponds to finding the function ${\cal F}$.  
Note that we need to make this border in high-dimensional feature space (although the present example in figure \ref{fig3}$(a)$ is an illustration for a low-dimensional classification problem).  
This mapping is determined by solving an optimization problem of
\begin{eqnarray}
{{\bm W} = {\rm argmin}_{\bm W}\biggl[\frac{1}{2}||{\bm W}||^2+C\sum^{N}_{i=1}h(y_i-{\cal F}({\bm x}_i; {\bm W}))\biggr]}\\
{\rm subject~to}~~~ h(y_i-{\cal F}({\bm x}_i))={\rm max}(0,|y_i-{\cal F}({\bm x}_i)|-e),
\end{eqnarray}
where $h$ is the $e$-insensitive loss function and $C$ is a coefficient of the error term, respectively. 
We are able to change the importance of each term by tuning the coefficient $C$.
The schematic of function {$h$} is shown in figure \ref{fig3}$(b)$.   
As shown here, the error $y-{\cal F}({\bm x})$ (horizontal axis) within the constraint called epsilon tube is judged to achieve an error of $h(y-{\cal F}({\bm x}))=0$ (vertical axis).  For instance, with the case that the output of machine learning model ${\cal F}(\bm x)$ shows slight difference from $y$ because of a noisy input, the error $h(y-{\cal F}({\bm x}))$ is converted to zero if the error value is within the constraint by the epsilon tube.  Thus, the epsilon tube can be regarded as the allowance against a noisy input.
We can tune the allowance (tolerance) against noise through the loss function by incorporating $e$ in the loss function (3).  Here, we optimize $C$ and $e$ through the optimization procedure outlined in section 2.1.1.

\subsubsection{Extreme learning machine}

Various machine learning models with reduced computational requirements have been proposed.  
Such techniques are often simplified against the original formulations by sparsification and randomization.  
These ideas are particularly attractive for fluid dynamics since the amount of training data is enormous in size.
In this paper, we consider the extreme learning machine (ELM) \cite{HZS2004}, which is a simplified version of the MLP.  
Its computational cost is significantly lower than the original MLP due to its simple construction.  
The basic structure of ELM is similar to MLP with only a single hidden layer.  
First, weights matrix between input and hidden layers with $n$ nodes ${\bm W}_1 = [{\bm w}_1^1,{\bm w}_2^1,\cdot\cdot\cdot, {\bm w}_n^1]^{\rm T}$ are initialized by non-zero random values.   
This weights matrix and activation function $\varphi$ are multiplied for input data matrix ${\bm q}_{\rm in}$ in a manner similar to the MLP.  

In ELM, the output data matrix of the first layer ${\bm q}_h$ is expressed as
\begin{eqnarray}
    {\bm q}_h = \varphi({\bm W}_1 {\bm q}_{\rm in}).
\end{eqnarray}
The weights matrix between hidden and output layers ${\bm W}_2 = [{\bm w}_1^2,{\bm w}_2^2,\cdot\cdot\cdot, {\bm w}_n^2]^{\rm T}$ are determined through the least-square formulation (pseudoinverse) such that
\begin{equation}
    {\bm W}_2 = {\bm q}_h^* {\bm q}_{\rm out}^T,
\end{equation}
where {$*$ denotes the Moore--Penrose pseudoinverse} \cite{Albert1972,Serre2002} and ${\bm q}_{\rm out}$ is the output data matrix of the last layer.   Only the number of units is optimized in this paper.  We refer readers to Maulik and San \cite{MS2017} for additional details. 

\subsubsection{Convolutional neural network}

In recent years, convolutional neural networks (CNN) \cite{LBBH1998} have become heavily utilized, especially for image recognition tasks.  
One of the attractive functions of CNNs is their ability to handle high-dimensional data without encountering the curse of dimensionality, by incorporating filtering operations.  
Such capability has led to the increased uses of CNN, including the field of fluid dynamics \cite{SP2019,FFT2019,FNKF2019}.

\begin{figure}
	\vspace{0mm}
	\begin{center}
		\includegraphics[width=0.65\textwidth]{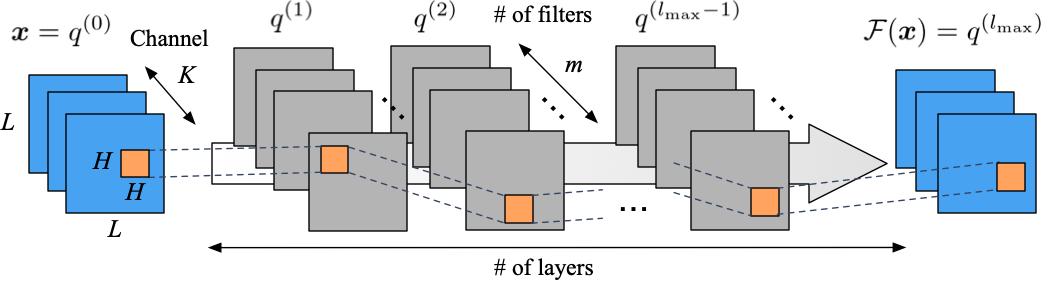}
		\caption{Two-dimensional convolutional neural network.}
		\label{fig4}
	\end{center}
\end{figure}

The schematic of two-dimensional CNN is illustrated in figure \ref{fig4}.  
Now, let us consider an input data ${\bm x}=q^{(0)}$ which has $L\times L$ positions.  
The process of CNN can be expressed in a procedural manner from $q^{(l-1)}$ to $q^{(l)}$, where $l~(0\leq l \leq l_{\rm max})$ represents the index of the network layer.  
The mathematical expression for $q^{(l)}$ is given as 
\begin{eqnarray}
    q^{(l)}_{ijm} = {\varphi}\biggl(\sum^{K-1}_{k=0}\sum^{L-1}_{p=0}\sum^{L-1}_{s=0}h_{p{s}km}^{(l)} q_{i+p,j+{s},k}^{(l-1)}\biggr),
\end{eqnarray}
where $q^{(l_\text{max})} = {\mathcal F}({\bm x})$ and $K$ is the number of variables per each position of data. 
The orange $H\times H$ square in the schematic represents the filter $h$. 
The CNN can be considered as a framework to optimize the filtering coefficients in the above equation, which is analogous to the weights of MLPs as formulated as (\ref{eq:1}).  
The concept of filters in CNN is called weight sharing because the filter of size $H\times H$ is shared for whole image field of size $L\times L$.  
By using this weight sharing instead of the fully-connected weights in the MLP, we can handle big data with significantly lower computational costs.  
In the current work, three-dimensional CNN (with a three-dimensional filter of size $H\times H \times H$) is also utilized for the example of super-resolution reconstruction.  Moreover, we utilize a customized CNN called {hybrid downsampled skip-connection and multi-scale (DSC/MS) model} for super-resolution analysis, which will be described in detail later.


\subsection{Choice of norms in loss function}
\label{sec:norm}

\begin{table}
\begin{center}
\caption{Commonly used loss functions for regression tasks.}
\def~{\hphantom{0}}
\begin{tabular}{cccc}
     \hline \hline
        Error & $L_1$ error & $L_2$ error & $L_2$ logarithmic error\\ \vspace{1mm}
        Definition & $||{\bm q}_{\rm Ref}-{\bm q}_{\rm ML}||$ & $||{\bm q}_{\rm Ref}-{\bm q}_{\rm ML}||_2$ & $||{\rm log}(1+{\bm q}_{\rm Ref})-{\rm log}(1+{\bm q}_{\rm ML})||_2$\\ \vspace{1mm}
        Note & Not sensitive for an outlier & Strict for an outlier & Tends to overestimate the value\\
        \hline \hline
    \end{tabular}
  \label{tab2}
\end{center}
\end{table}

Let us discuss briefly the influence of loss functions on the performance of machine learning models.  
During the training process, supervised machine learning models aim to minimize the error between their output ${\cal F}({\bm x})$ and an answer $\bm y$.
Therefore, the choice of norm influences the updating procedure for the weights of machine learning models.
We list commonly used loss functions for regression problems in table \ref{tab2}.   
In the present study, we use the $L_2$ norm error.  
Due to its definition including squared procedure, squared error is the strict measurement including outliers.  
On the other hand, the $L_1$ norm error is not as sensitive for outliers in the data set.  
The squared logarithmic error is generally utilized for cases where underestimation should be avoided (e.g., prediction of the number of customers for a restaurant, in which case overestimation is preferred).  
Alternatively, we can also incorporate the regularization term into a loss function to take a constraint for weights \cite{Bishop2006}.  
It is widely known that the use of the regularization for weights helps machine learning models attain robustness for overfitting and have weights to be a sparse.  
Depending on the problem setting and characteristics of the provided data sets, one can choose the appropriate loss function.      

\subsection{Cross validation and overfitting}
\label{sec:cv}

\begin{figure}
	\vspace{0mm}
	\begin{center}
		\includegraphics[width=0.80\textwidth]{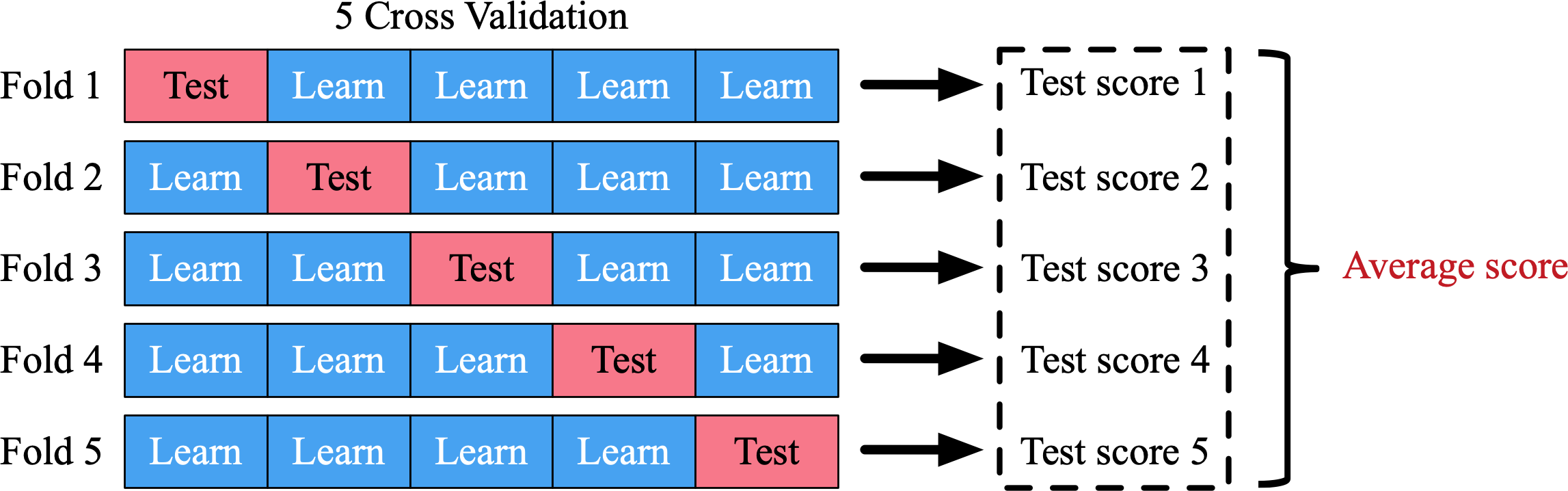}
		\caption{
		An example of five-fold cross validation.  The blue boxes indicate training data used for deriving machine learning model.  The red boxes indicate test data used to obtain test scores.  An average score of five test scores are used for the optimization procedure and assessment.
		}
		\label{figA1}
	\end{center}
\end{figure}

Presented in figure \ref{figA1} is the five-fold cross validation used in the present study.  
In this example, the data set is divided to five subsets.  
To maintain reproducibility, we should use different boxes for each cross validation process as the test data set colored by red in figure \ref{figA1}.  
We then obtain an average score from five for assessment.  
Supervised machine learning model should be able to provide good predictions for unseen data from this training process, thus avoiding overfitting of data. 
This process is critically important as the quote by Brunton and Kutz \cite{BK2019} suggests: ``{\it if you don't cross-validate, you is dumb.}''

Furthermore, we can use various functions to prevent overfitting, i.e., dropout \cite{SHKSS2014}, $L_1$ or $L_2$ regularizations of loss function \cite{Bishop2006}.  
However, we do not explore these options in the present study for brevity.


\section{Examples}
\label{sec:examples}

For demonstration, we consider three regression problems in fluid dynamics (as summarized in figure \ref{fig00}):
\begin{enumerate}
\item The estimation of drag and lift coefficients from limited sensors in laminar flows; 
\item The estimation of laminar wake flow from limited sensors; and
\item Super-resolution analysis of coarse images for laminar and turbulent flows
\end{enumerate}
which are listed in the order of complexity.
To tackle these problems, the following four machine learning models are tested:
\begin{enumerate}
\item Multi-layer perceptron (MLP) (section 2.1.1)
\item Random forest (RF) (section 2.1.2)
\item Support vector regression (SVR) (section 2.1.3)
\item Extreme learning machine (ELM) (section 2.1.4)
\end{enumerate}
for Problems 1 and 2.  For Problem 3, we use:
\begin{enumerate}
\item Two- and three-dimensional convolutional neural networks, and
\item Two- and three-dimensional DSC/MS model.
\end{enumerate}

{
We note that normalization and standardization are not applied to the current examples.
Readers can also consider the use of these scaling methods to improve the quality of approximations \cite{SHH1996}.
}

\subsection{Estimation of drag and lift coefficients for laminar wakes}

\begin{figure}
	\vspace{0mm}
	\begin{center}
		\includegraphics[width=1.00\textwidth]{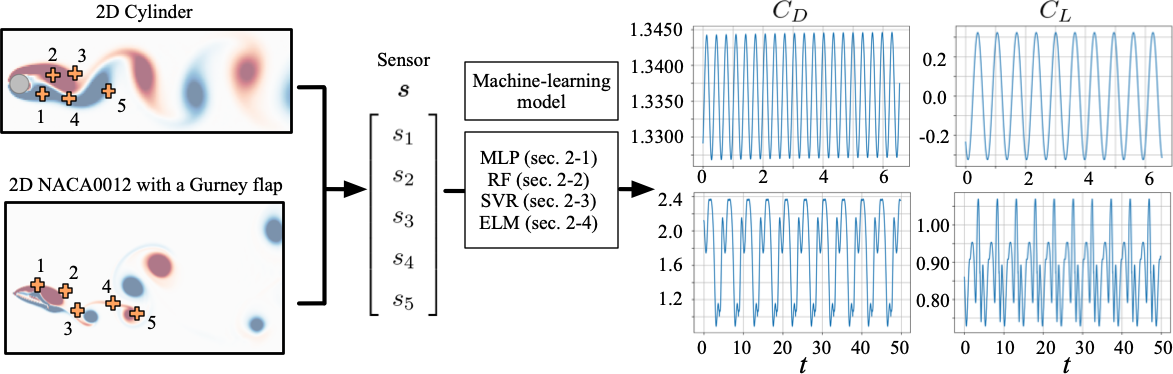}
		\caption{
		Estimation of $C_D$ and $C_L$ values with laminar flow.  Models for estimating $C_D$ and $C_L$ are constructed individually.}
		\label{fig5}
	\end{center}
\end{figure}

Various machine learning methods have been used to develop a suitable meta-modeling architecture for various flows.  
In earlier studies, there were efforts to estimate drag or lift coefficient from flow field data.
Zhang {\it et al}.~\cite{ZSM2018} tried to estimate the lift coefficient from flow characteristics and geometry of NACA632615 airfoil.  
One of the interesting observations of their study is that the test accuracy can be improved by using an artificial image as the input, which has a color corresponding to the input parameters.
The estimation method for drag and lift coefficient of various bluff body shapes were proposed by Miyanawala and Jaiman \cite{MJ2018}.  
They achieved 80\% accuracy for test data by converting the regression problem into a classification problem.  
Here, we focus on the drag and lift coefficient estimations for the wakes behind a two-dimensional cylinder and NACA0012 airfoil with a Gurney flap.  
In the present setting, we only use five sensor measurements in the wake, as illustrated in figure \ref{fig5}.  
We examine this simple problem to assess the performance of the four machine learning models.

Let us consider the two-dimensional cylinder wake flow. 
The flow fields are prepared from a two-dimensional direct numerical simulation \cite{TC2007,CT2008}. 
The governing equations are the incompressible Navier--Stokes equations:
\begin{eqnarray}
{\bm\nabla} \cdot {\bm u}= 0, \\
\frac{\partial {\bm u}}{\partial t}+{\bm u} \cdot {\bm\nabla} {\bm u}=-{\bm\nabla} p+\frac{1}{Re_D}\nabla^2\bm u,
\end{eqnarray}
where $\bm u$ and $p$ are the non-dimensional velocity and pressure variables.  
The Reynolds number $Re_D=100$ is based on the diameter and the freestream velocity.
Five nested levels of multi-domains are used with the finest level being $(x, y)/D = [-1, 15] \times [-8, 18]$ and the largest domain being $(x,y)/D = [-5, 75] \times [-40, 40]$ in size.
The time step is set to $\Delta t=2.50\times 10^{-3}$.
In example 3.1, we utilize the data around the cylinder body: the size of domain of interest and the number of grid points set to $[-0.7, 15] \times [-5, 5]$ and $(N_x^*, N_y^*) = (192, 112)$, respectively.

\begin{figure}
	\vspace{0mm}
	\begin{center}
		\includegraphics[width=1.00\textwidth]{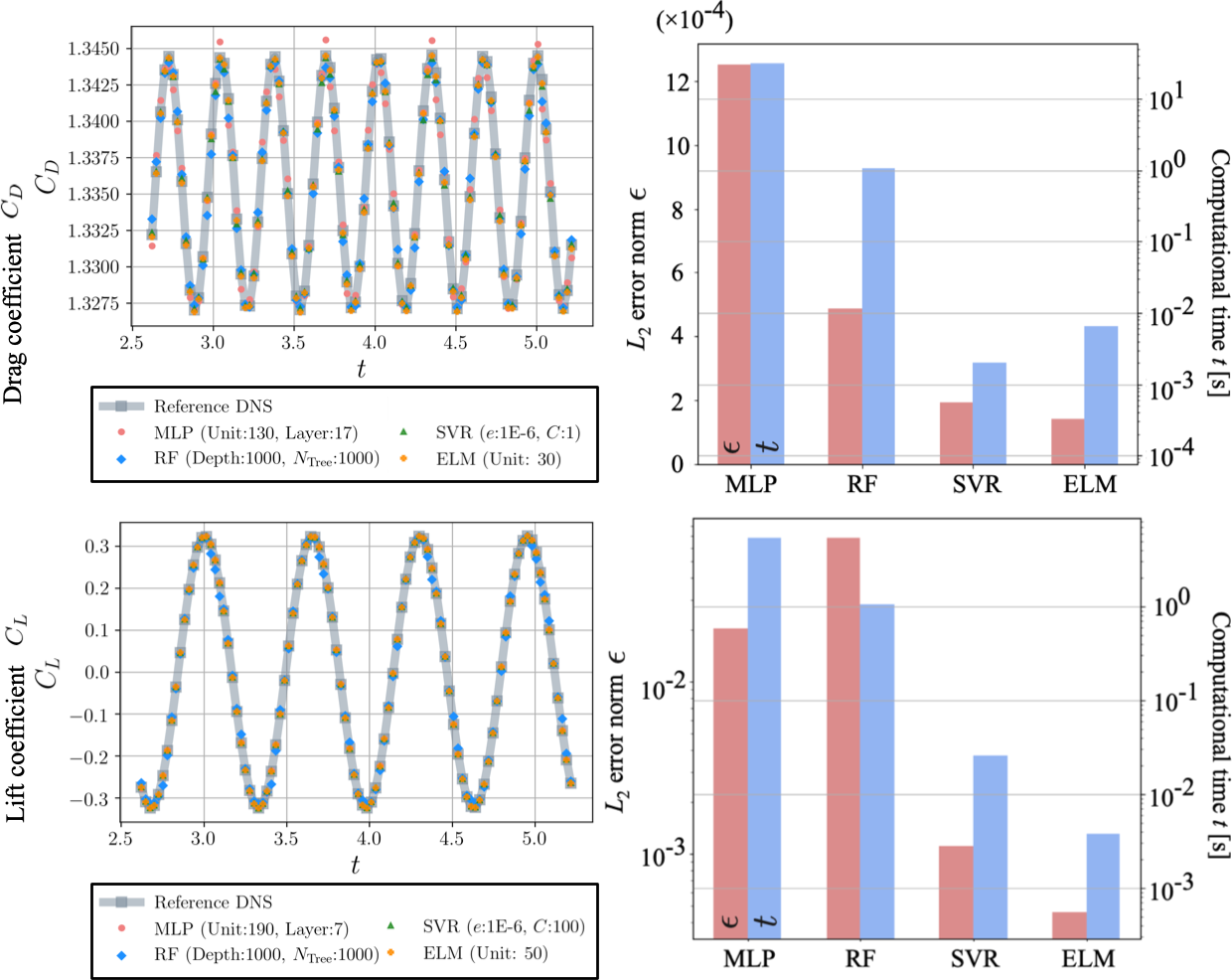}
		\caption{
		Summary of $C_D$ and $C_L$ estimations for laminar cylinder wake at $Re_D=100$.  In the top and bottom, $C_D$ and $C_L$ estimations, respectively, are presented.  The left side shows temporal variations of force coefficients.  The optimized parameters are provided in the graph legends.  The color bars on the right show the relationship between $L_2$ error norm $\epsilon$ of each coefficient (red) and computational time (blue) for each machine learning model.}
		\label{fig6}
	\end{center}
\end{figure}

As shown in figure \ref{fig5}, we consider readings from the five sensor measurements ${\bm s}=\{s_1,s_2,s_3,s_4,s_5\}$ in the wake as the input data to the machine learning models in order to estimate the force coefficients through 
\[
C_D={\mathcal F}_D(\bm s)
\quad \text{and} \quad
C_L={\mathcal F}_L(\bm s).
\] 
Here, the vorticity value $\omega$ is used as the input attribute.  Note here that we make a machine learning model for each coefficient separately {to avoid the influence of the magnitude differences on the model}.
The comparison of the results from the four machine learning models are summarized in figure \ref{fig6}.
We use 100 snapshots corresponding to 8 periods in time as the training data for the baseline model.  
The top and bottom subplots represent the $C_D$ and $C_L$ estimations, respectively.  
The plots in the left side present the time history of force coefficients.
As mentioned above, the parameters of each machine learning model are optimized for fair comparison and shown in the graph legend.  
For both $C_D$ and $C_L$ estimations with all machine learning models, the results well with the reference data. 
The assessment of $L_2$ error norm $\epsilon = ||f_{{\rm True}}-f_{{\rm Pred}}||_2/||f_{{\rm True}}||_2$ and the computational time {for the learning process} on the NVIDIA Tesla K40 graphics processing unit (GPU) for the four machine learning models are shown on the right side of figure \ref{fig6}.  
The errors are approximately $0.2\%$ for $C_D$ and $1\%$ for $C_L$ for the all machine-learned models, respectively.  
Note that these errors and computational costs are assessed with five-fold cross validation.  
With both force estimations, the ELM has advantage over other machine learning models in terms of the accuracy and computational cost.  
Note that although the ELM shows advantage in terms of both accuracy and computational cost, this method is not robust against noisy inputs due to its simple structure, as discussed later.  
{ Also, the comparison here is not straightforward for the computational costs since the parameters of each model is optimized for accuracy.}
Overall, all machine learning models work as expected since this problem is simple due to the periodic nature of the cylinder wake.

Next, let us consider another example of complex two-dimensional wake behind a NACA0012 airfoil with a Gurney flap as shown in figure \ref{fig5}.  
Although the flow field is periodic in time, the problem setting here can be regarded as tougher than the cylinder problem because the wake behind an airfoil with a Gurney flap has various frequency contents as explained later.
The training data set is obtained by two-dimensional direct numerical simulation at $Re_c =1000$, defined based on the chord length $c$ \cite{GMTA2018}.  
Various types of wakes can be generated depending on the angle of attack $\alpha$ and the Gurney-flap height $h/c$.  
In what follows, we focus on the case of $h/c=0.1$ and $\alpha = 20^\circ$ which produces a complex unsteady wake comprised of time-periodic shedding of two individual vortex pairs, as shown in figure \ref{fig5}. 
As a signature of two vortex pairs shedding from the airfoil, we observe two dominant frequencies \cite{GMTA2018}.
For this problem, we use the five nested levels of multi domains.  
The finest domain range is ($x/c$, $y/c$) = $[-1, 1] \times [-1, 1]$ and the largest domain is ($x/c$, $y/c$) = $[-16, 16] \times [-16, 16]$.  
The time step is set to $\Delta t=10^{-3}$.
In example 3.1, we use the data around the airfoil: the size of utilized domain and the number of grid points are $[-0.5, 7] \times [-2.5, 2.5]$ and $(N_x^*, N_y^*) = (352, 240)$, respectively.  
Analogous to the cylinder example, we use the vorticity field $\omega$ as the input attribute to the machine learning models.  

\begin{figure}
	\vspace{0mm}
	\begin{center}
		\includegraphics[width=1.00\textwidth]{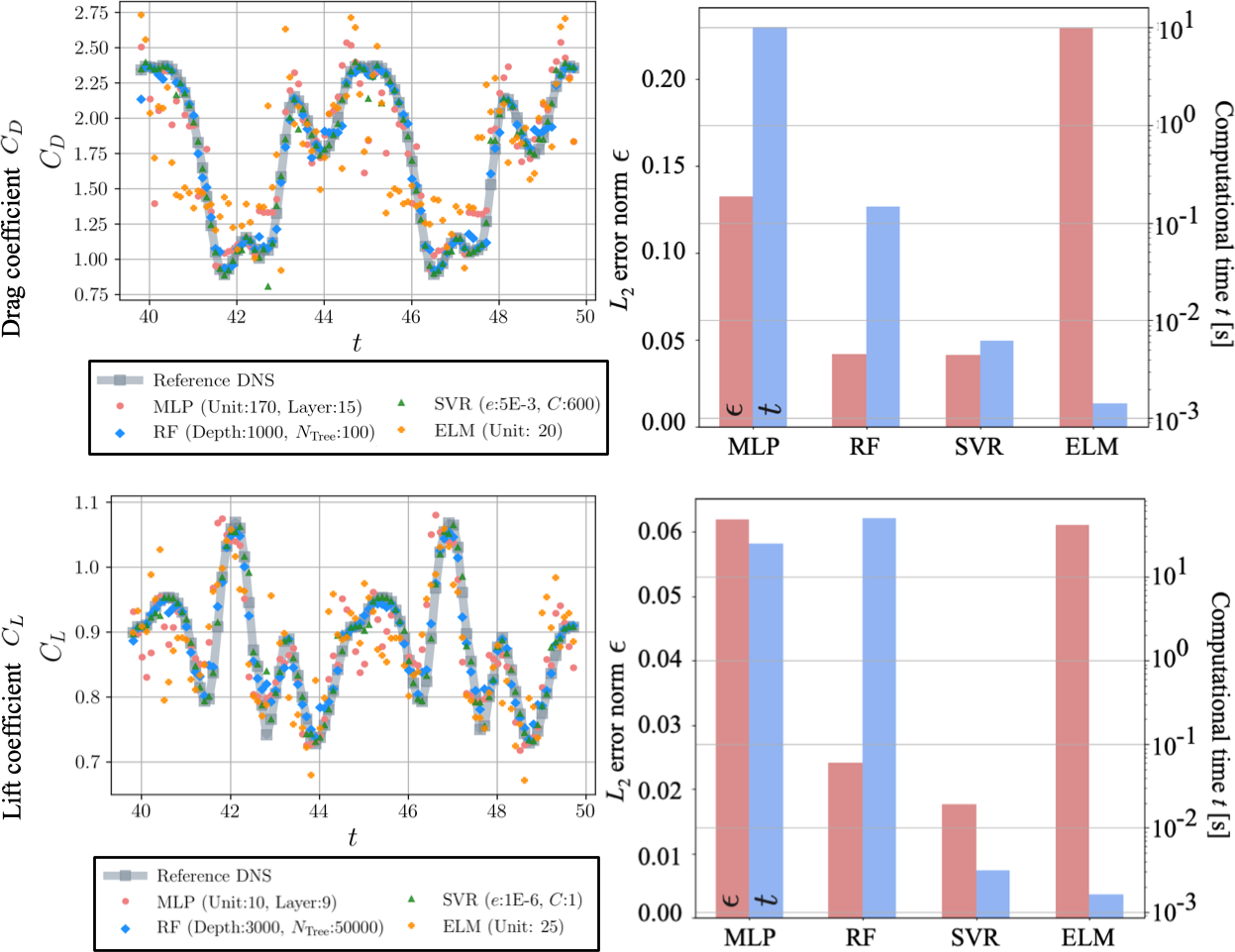}
		\caption{
		Summary of $C_D$ and $C_L$ estimations for wakes of NACA0012 airfoil with a Gurney flap at $Re_c=1000$.  The top left shows $C_D$ estimations and bottom left presents $C_L$ estimations.  The optimized parameters are provided in the graph legends.  The color bars located on the right show the relationship between $L_2$ error norm $\epsilon$ of each coefficient (red) and computational time (blue) for each machine learning model.}
		\label{fig7}
	\end{center}
\end{figure}

\begin{figure}
	\vspace{0mm}
	\begin{center}
		\includegraphics[width=0.90\textwidth]{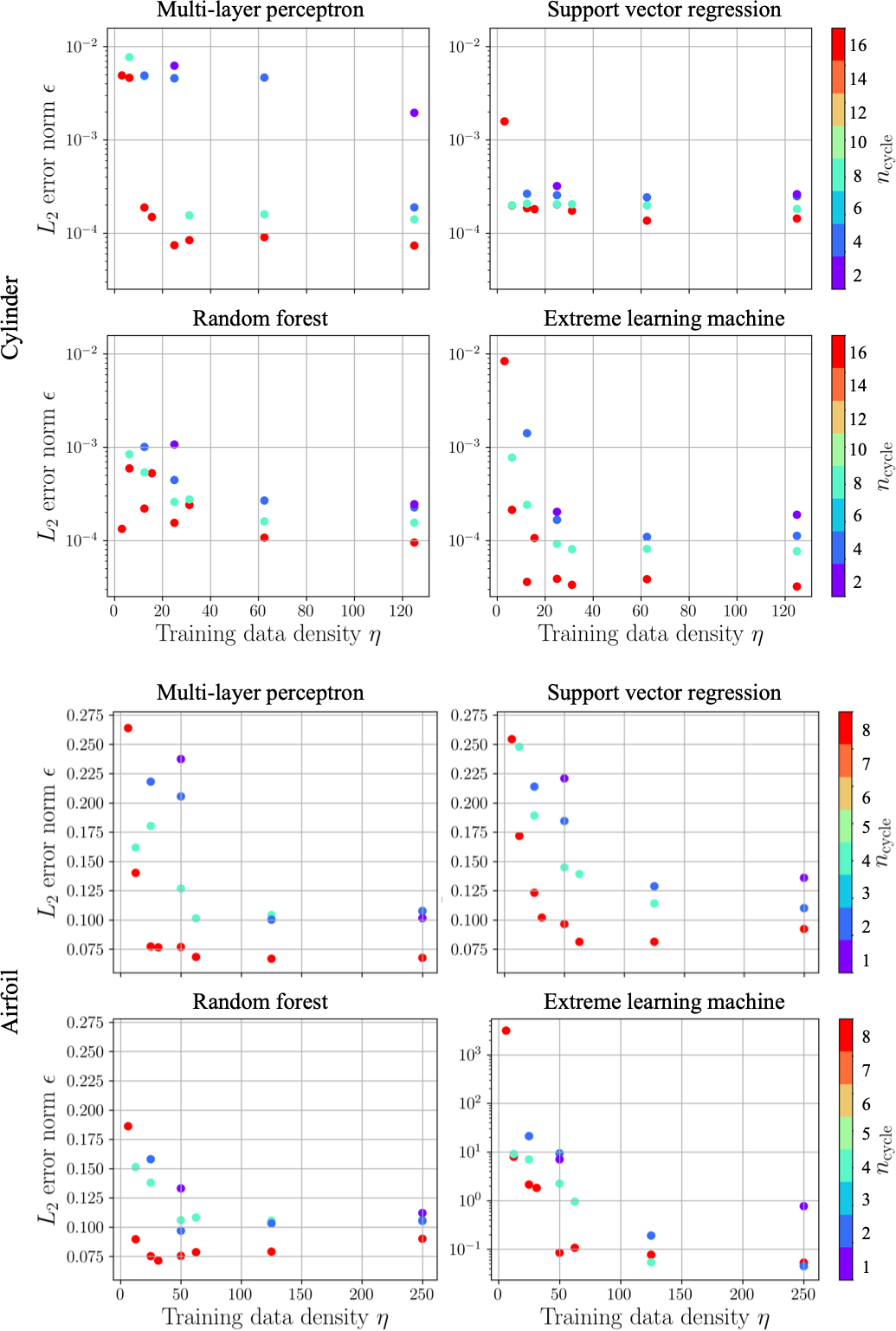}
		\caption{The relationship between the training data density $\eta$ and $L_2$ error norm $\epsilon$ of $C_D$ prediction for four machine learning models.  Note that the range of $y$-axis with extreme learning machine for the airfoil problem is different from the others.}
		\label{fig8}
	\end{center}
\end{figure}

The summary of $C_D$ and $C_L$ estimations for flows over a NACA0012 airfoil with a Gurney flap is shown in figure \ref{fig7}.  
We use 100 snapshots corresponding to 8 periods in time as the training data following the cylinder problem.  
While this problem setting is somewhat more complicated than the cylinder example, the force estimations from MLP, SVR, and RF are in agreement with the references solution.  
However, the error from the ELM is notably high for this problem setting, although this model exhibits great performance for the cylinder problem.  
It suggests that an ELM is not necessarily robust for complex problems.  
Note that the ELM is able to accurately estimate the forces well when a large number of snapshots are available for training, as explained later with figure \ref{fig8}.  
When complexity is added to the flow problems, the accuracy of the models is compromised especially for those with lower computational costs. 
This point should be carefully considered when one chooses a machine learning model.  
Additional computational cost is likely necessary for robustness, as this example suggests.

The cylinder wake and NACA0012 flows are periodic in time.  
The accuracy of machine-learned models depends on the number of snapshots per cycle (temporal period) contained in the training data. 
The relationship between the training data density $\eta$ and $L_2$ error norm $\epsilon$ of $C_D$ estimation with laminar problems are summarized in figure \ref{fig8}.  
Here, the training data density $\eta$ is defined as $\eta \equiv n_{\rm snapshot}/{n_{\rm cycle}}$, where $n_{\rm snapshot}$ is the number of snapshots and $n_{\rm cycle}$ is the number of cycles in time, respectively.  
The tuning parameters of the machine learning models are the same as those in the above discussions (obtained from $n_{\rm snapshot}=100$ and $n_{\rm cycle}=8$). 
In general, higher density leads to enhanced estimation.  
Noteworthy here is that increased number of cycles for the training data improves model accuracy.  
Although repeating data sets would not add merits to the training process, slight offset in phase can provide additional merits, which likely happened here.
The other notable observation is that the ELM is not robust for short training data for the airfoil problem.  
Although the computational time to establish the ELM is much shorter compared to other models, it requires larger amount of data for reconstructing problems with added complexity.  
We note in passing that similar trends are observed with $C_L$ estimations.

\begin{figure}
	\vspace{0mm}
	\begin{center}
		\includegraphics[width=1.00\textwidth]{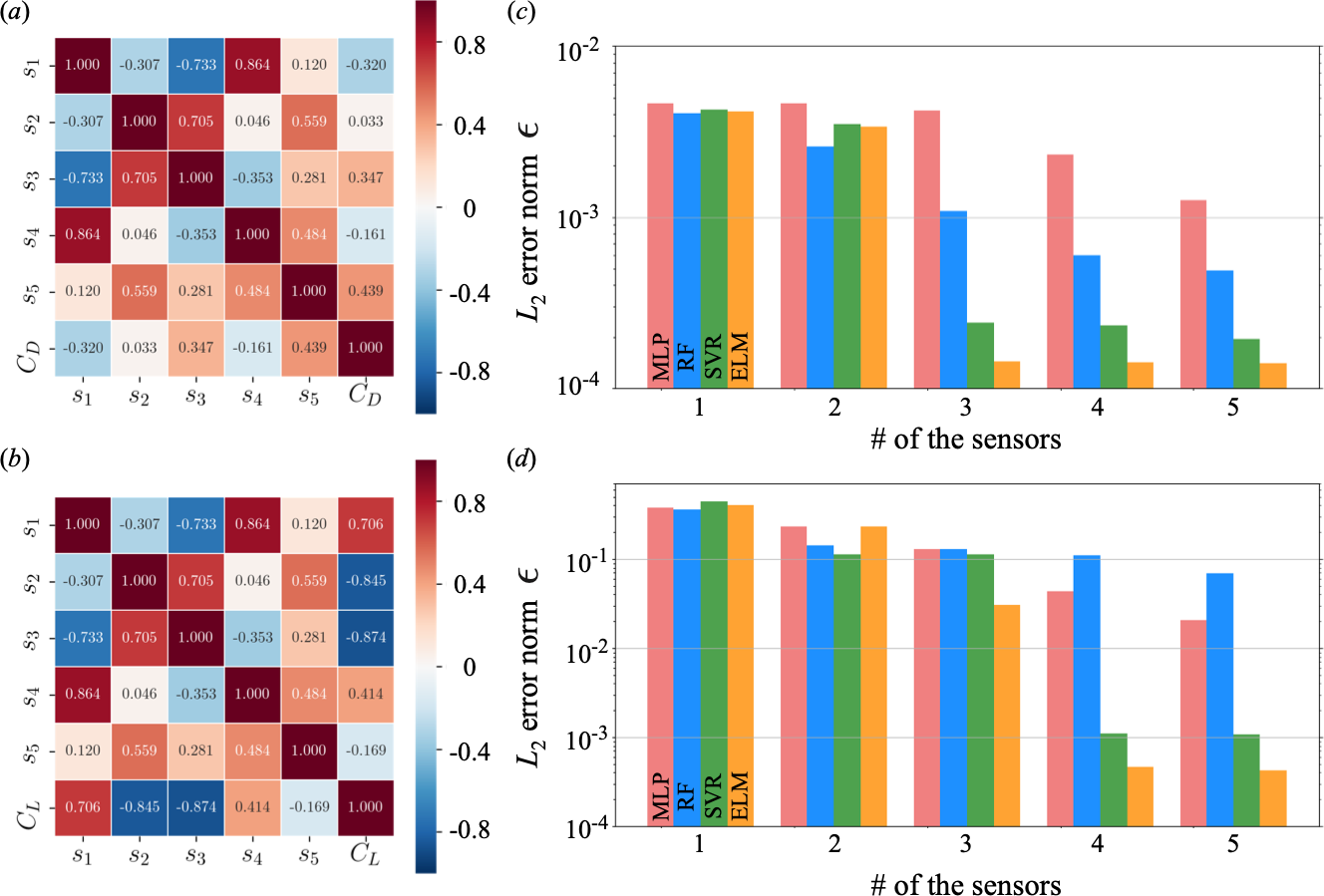}
		\caption{The influence of the number of sensors $n_{\rm sensor}$ on the accuracy of the force estimations. Correlation matrix of $(a)$ $C_D$ and $(b)$ $C_L$ estimations. The relationship between $n_{\rm sensor}$ and $L_2$ error norm of $(c)$ $C_D$ and $(d)$ $C_L$ estimations.}
		\label{fig12_20191230}
	\end{center}
\end{figure}

{
In the above discussions, the five sensor measurements are used as the input for the machine-learning models.
Next, let us investigate the influence of the number of sensors $n_{\rm sensor}$ for the force estimations.
We examine $n_{\rm sensor}=\{1,2,3,4,5\}$.
The number of sensors are reduced based on the magnitude of correlation coefficients with other sensors and force coefficients, as explained in detail later.
The normalized correlation coefficient is given as
\begin{equation}
    \large
    R_{ij} = \frac {C_{ij}}{\sqrt{C_{ii}\cdot C_{jj}}},
\end{equation}
where $C_{ij}$, $C_{ii}$, and $C_{jj}$ denote the covariance matrices corresponding to that of a single sensor and its counterpart, the single sensor itself, and its counterparts themselves, respectively. 
The dependence on $n_{\rm sensor}$ with the covariance matrices for both $C_D$ and $C_L$ estimations are shown in figure \ref{fig12_20191230}.
Since the correlation magnitude of $s_2$ is the lowest for $C_D$, we use ${\bm s}_{\rm reduced}=\{s_1,s_3,s_4,s_5\}$ with $n_{\rm sensor}=4$.
In other words, the sensor with the lowest correlation magnitude is discarded with decreasing the number of used sensors as the input.
Note that the optimized parameter at $n_{\rm sensor}=5$ is used for construction of machine learning models with other number of the sensors.
With both force estimations, the errors are dramatically increased when the high-correlation sensors are removed, i.e., $n_{\rm sensor}=2$ for $C_D$ estimation (when $s_1$ is removed) and $n_{\rm sensor}=3$ for $C_L$ estimation (when $s_4$ is removed).
Although we apply the correlation analysis in the case of this example, which were randomly distributed as in figure 8, other techniques such as $K$-means clustering \cite{MacQueen1967} can be considered alternatively for the initial problem setting.
}

\subsection{Estimation of laminar wakes from limited sensor measurements}

\begin{figure}
	\vspace{0mm}
	\begin{center}
		\includegraphics[width=1.0\textwidth]{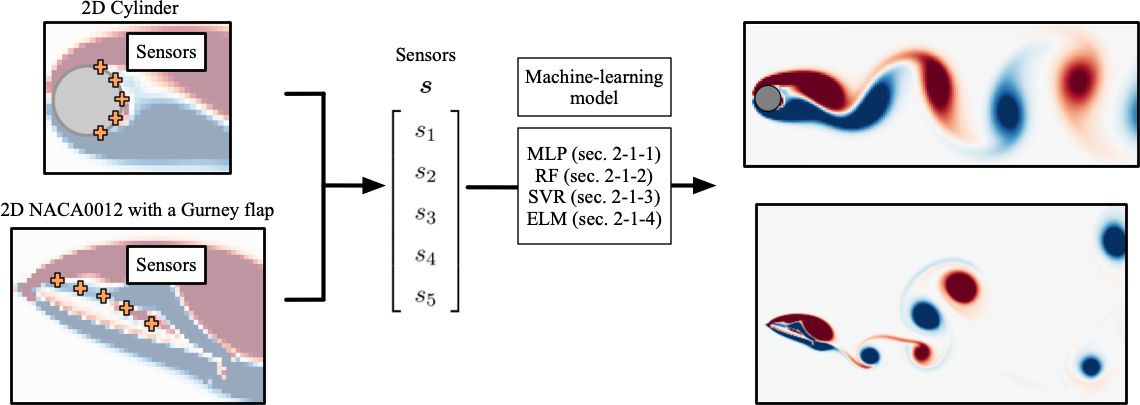}
		\caption{
    	Wake reconstruction using machine learning with five surface-sensor measurements.}
		\label{fig9}
	\end{center}
\end{figure}

Next, let us significantly increase the dimension and complexity of the problem. 
Here, we consider reconstructing the wake from limited measurements, as illustrated in figure \ref{fig9}.  
This problem settings for machine learning assessments are inspired by Erichson {\it et al.} \cite{EMYBMK2019}. 
They reconstructed the flow field from limited measurements using MLP referred to as the {\it shallow decoder}.  
We perform the assessments for cylinder flows and the NACA0012 wakes.  
In both cases, we use 100 snapshots corresponding to 8 periods as the training data.  
To focus on the wake region, we extract the flow field from the original DNS data $(N_x^*, N_y^*) = (192, 70)$ with $(x^*,y^*)/D = [-0.7,15] \times [-3.3,3.3]$ and $(N_x^*, N_y^*) = (352, 192)$ with $(x^*,y^*)/c = [-0.5,7]\times [-2,2]$ respectively for the cylinder and NACA0012 flows.

\begin{figure}
	\vspace{0mm}
	\begin{center}
		\includegraphics[width=1.0\textwidth]{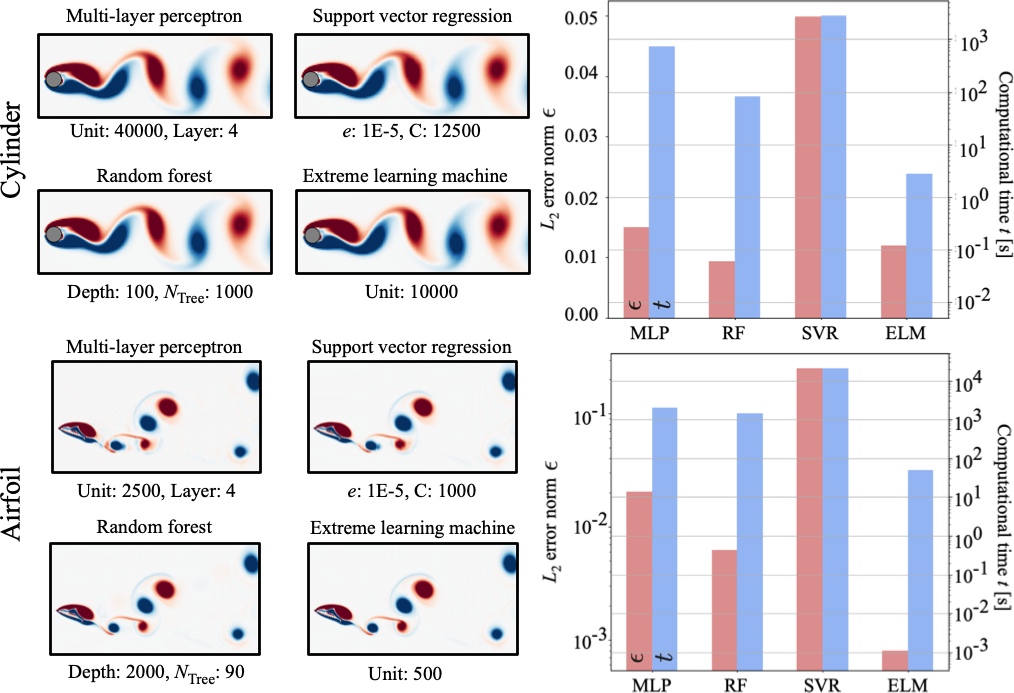}
		\caption{
		Summary of wake reconstruction for cylinder and NACA0012 with Gurney flap wakes.  The left side shows the compilation of vorticity fields reconstructed by the four machine learning models.  Optimal parameters are listed below the vorticity fields.   The color bars located on the right side show the relationship between $L_2$ error norm $\epsilon$ (red) and computational time (blue) for the selected machine learning models.}
		\label{fig10}
	\end{center}
\end{figure}

The results of the cylinder wake reconstruction are summarized in the top of figure \ref{fig10}. 
For the cylinder case, the machine learning models $\cal F$ are the mapping functions from five sensor measurements $\bm s \in {\mathbb R}^5$ to the whole wake flow field ${\cal F}({\bm s}) \in {\mathbb R}^{13440}$ such that {${\bm \omega}\approx{\cal F}(\bm s)$}.  
The vorticity field $\omega$ is used as the input and output attributes.  
The flow field can be reconstructed well with all machine learning models as shown by the contour plots in figure \ref{fig10}.  
The relationship between $L_2$ error norm $\epsilon$ and computational costs are also evaluated.  
It can be seen that the SVR is {less} suitable for this high-dimensional output problem {because the error levels are higher than those of the other models despite its higher computational costs}.  

We also consider the reconstruction of the wake of a NACA0012 airfoil with a Gurney flap, as summarized in the bottom of figure \ref{fig10}.  
With the airfoil wake, the machine learning models attempt to reconstruct the vorticity field ($\in {\mathbb R}^{67584}$) from five sensor measurements located on the upper surface.  
The reconstructed flow fields show reasonable agreement with the reference DNS data as shown in figures \ref{fig9} and \ref{fig10}, although the $L_2$ error norms are slightly larger than those from the cylinder example {except for ELM} due to the increased complexity of the wake.  
In terms of the errors and computational costs, the trends for this problem follow the cylinder wake example.

\begin{figure}
	\vspace{0mm}
	\begin{center}
		\includegraphics[width=1.0\textwidth]{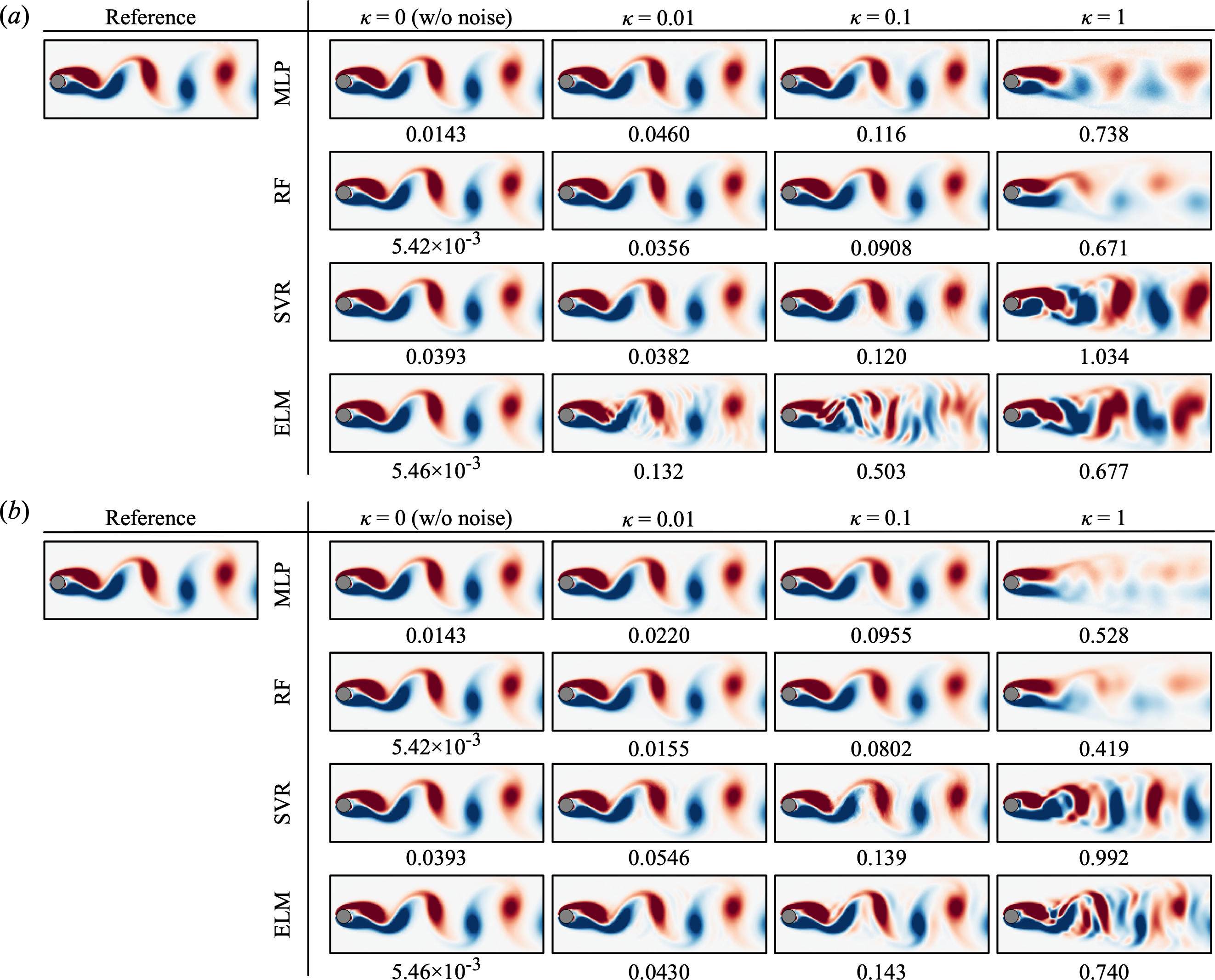}
		\caption{
		The reconstructed vorticity fields of cylinder wake from five noisy surface-sensor measurements $({\bm s}+\kappa {\bm n})$.  The values listed underneath each flow field represent the $L_2$ error norms.  $(a)$ Without noise addition and $(b)$ with noise addition in training data.}
		\label{fig11}
	\end{center}
\end{figure}

\begin{figure}
	\vspace{0mm}
	\begin{center}
		\includegraphics[width=0.9\textwidth]{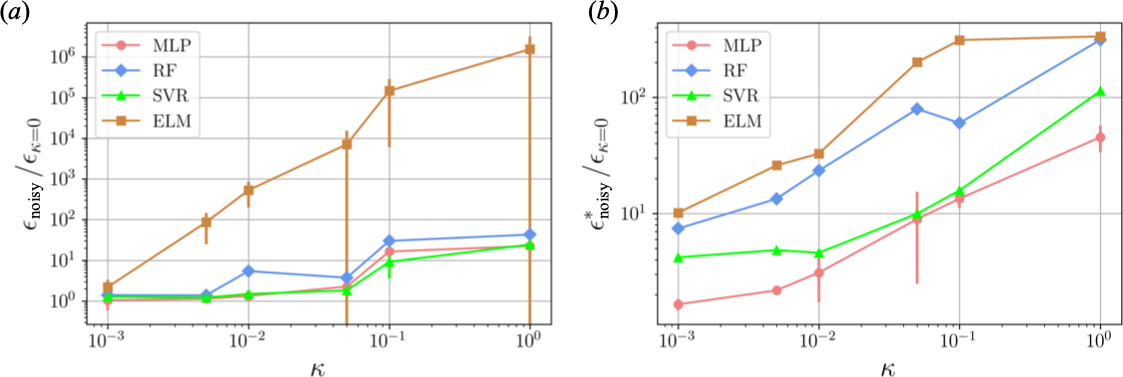}
		\caption{
		The $L_2$ error norms of the four machine-learned models over varied noise level $\kappa$.  The error levels are normalized by the error level without noise added to training data.  $(a)$ Without noise addition and $(b)$ with noise addition in training data.}
		\label{fig12}
	\end{center}
\end{figure}

Next, let us consider the robustness of these machine learning models against a noisy input by using the example of the wake reconstruction from the sensor measurements for each machine learning model.
Here, we consider two cases to examine the robustness as follows:
\begin{enumerate}
\item The machine-learned models {pre-trained} without noise are used as they are, while noise is added to the input data. The $L_2$ error norm of noisy input case is assessed as $\epsilon_{\rm noisy}=||{\bm \omega}_{\rm True}-{\cal F}({\bm s}+\kappa {\bm n})||_2/||{\bm \omega}_{\rm True}||_2$, where $\bm n$ is the Gaussian noise and $\kappa$ represents the magnitude of the noise.
\item Machine learning models are newly trained with training data containing noise such that $\epsilon^*_{\rm noisy}=||{\bm \omega}_{\rm True}-{\cal F}^*({\bm s}+\kappa {\bm n})||_2/||{\bm \omega}_{\rm True}||_2$, where ${\cal F}^*$ is machine learning models trained by using the training data with added noise.
\end{enumerate}
The reconstructed vorticity fields of the cylinder wake from five noisy sensor measurements are shown in figure \ref{fig11}.  
The value shown below each contour is the $L_2$ error norm in the respective case.
The relationship between the intensity of noisy inputs and $L_2$ error norms are also summarized in figure \ref{fig12}.  
In the cylinder flow case with MLP, RF, and SVR, the reconstructed flow fields show reasonable agreement with the reference DNS data up to $\kappa=0.1$.  
Once $\kappa$ reaches $\kappa=1$, these three machine-learned models cannot reconstruct the wake well and $L_2$ errors become larger than $40\%$, as shown in figure \ref{fig11}.  
On the other hand, the vorticity field obtained by the ELM cannot be reconstructed well even $\kappa=0.01$, although the accuracy is recovered by adding the noise into the training data as shown in figure \ref{fig12}$(b)$.  
This trends can be observed especially for $\kappa=0.1$ in figure \ref{fig11}.  
ELM lacks robustness against for noisy input in both cases due to its simplicity.  
Of course, this simple network structure has advantage in terms of low computational cost as mentioned above.  
Note that similar trends are observed for the NACA0012 wake case.
Users are cautioned to choose the appropriate machine learning model based on the objective, robustness requirement, and available computational resources.

\begin{figure}
	\vspace{0mm}
	\begin{center}
		\includegraphics[width=0.9\textwidth]{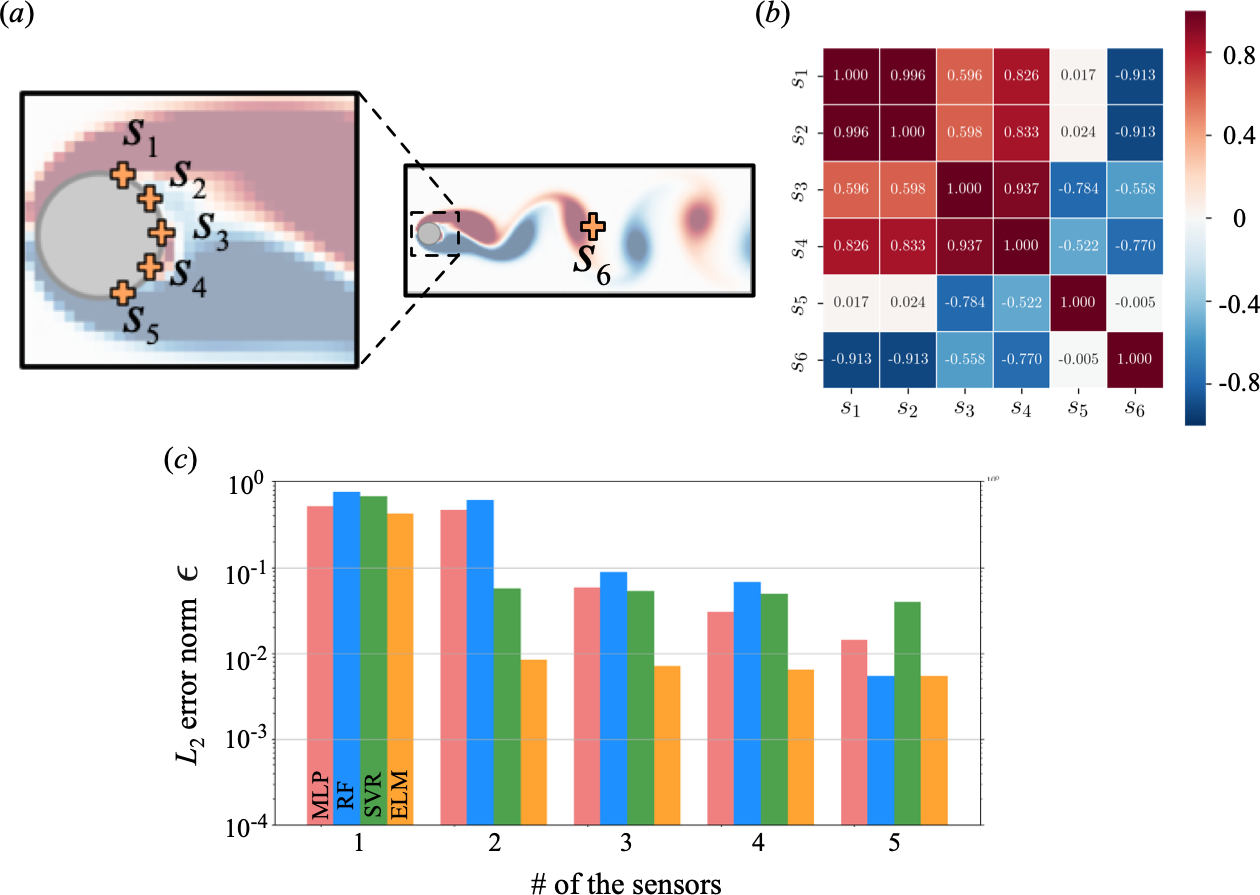}
		\caption{Influence of the number of sensors on the cylinder wake reconstruction. $(a)$ Positions of each sensor. $(b)$ Covariance matrix. $(c)$ Relationship between the number of sensors and $L_2$ error norm.}
		\label{fig17_20200105}
	\end{center}
\end{figure}

\begin{figure}
	\vspace{0mm}
	\begin{center}
		\includegraphics[width=1.0\textwidth]{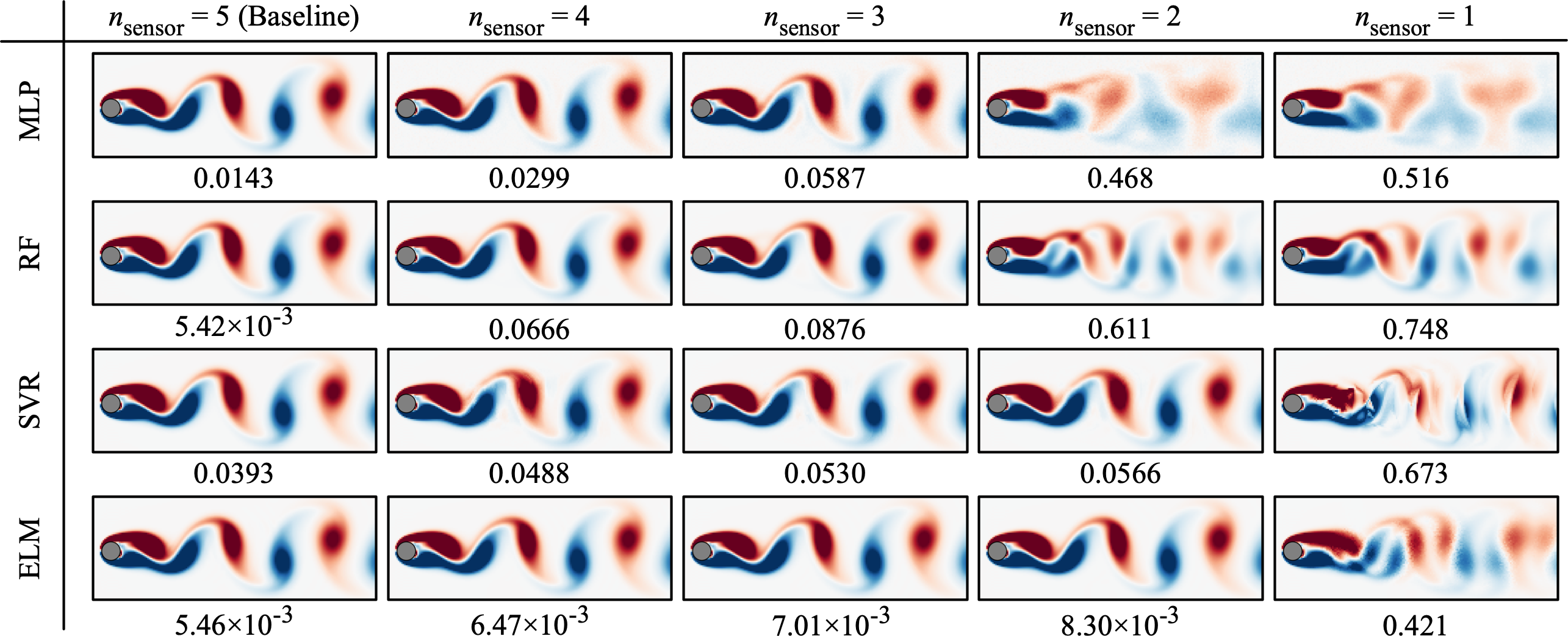}
		\caption{The reconstructed vorticity fields of cylinder wake with $n_{\rm sensor}=1,2,3,4,$ and 5.}
		\label{fig18_20200105}
	\end{center}
\end{figure}

{
The influence of the number of sensors $n_{\rm sensor}$ on the accuracy is investigated for the wake reconstruction using the cylinder example.
Instead of the force coefficients of $R_{ij}$ in section 3.1, we use a new sensor $s_6$ as the representation sensor of wake region as shown in figure \ref{fig17_20200105}$(a)$.
A covariance matrix is then constructed using ${\bm s}=\{s_1,s_2,s_3,s_4,s_5,s_6\}$.
We examine $n_{\rm sensor}=1,2,3,4,$ and 5.
Similar to the discussion of the influence on the number of sensors with the force estimation in section 3.1, the sensor with the low correlation magnitude is discarded with decreasing the number of used sensors as the input.
Note that the optimized parameter at $n_{\rm sensor}=5$ is used for construction of machine learning models with other number of the sensors.
As presented in figure \ref{fig17_20200105}$(c)$, the error increases with the reduction of the number of sensors.
The reconstructed wakes corresponding to each number of sensors with all machine-learned models are summarized in figure \ref{fig18_20200105}.
With $n_{\rm sensor}=3,4,$ and 5, the wakes are reconstructed well for all machine-learning models, although the errors increase slightly with lower $n_{\rm sensor}$.
The ELM and SVR can maintain the accuracy of reconstruction with $n_{\rm sensor}=2$.
On the other hand, the reconstructed fields with the MLP and RF do not capture the vortex shedding correctly.
For $n_{\rm sensor}=1$, all models are unable to reconstruct the wake with the present parameter settings of the machine learning models.
It implies that MLP and RF are more sensitive than SVR and ELM for the wake reconstruction problem in terms of parameter tuning.
Note here that the wake may be reconstructed with a small number of sensors and appropriate parameters, although we use the same parameters tuned for $n_{\rm sensor}=5$ in the present study.
}

\subsection{Super-resolution analysis of flow image data}

The curse of dimensionality is known as one of the biggest issues for machine learning in dealing with large data sets, such as images or videos \cite{Domingos2012}. 
For instance, applications of MLP is strongly restricted by this issue because of the fully-connected structure of MLP.  
In fact, four machine learning models discussed in section 3.2 cannot be easily applied to much higher dimensional problems as of this moment, due to the curse of dimensionality and the computational burden.  
To overcome this issue, the convolutional neural network (CNN) was developed in computer science \cite{LBBH1998}.  
The key feature of a CNN is called weight sharing which considers the weight as the filter operation as mentioned in section 2.1.5.  
The basic assumption of CNN is that the pixels of that are far apart in images have no strong relationship.
In fact, this assumption can be appropriate in many cases, allowing us to reconsider fully-connected structure.  
This is a very attractive idea for high-dimensional problems.
Recently, there have been increased attention on the use of weight sharing with fluid dynamics. 

One of the fields where this benefit is exploited is the example-based super resolution \cite{Bannore2009,Salvador2016} in computer science. 
Through the use of super-resolution reconstruction, a high-resolution signal can be reconstructed from a given low-resolution input.  
Although the concept here has been explored with various techniques including interpolations and high-frequency transfer, supervised machine learning has recently made tremendous strides with the super-resolution analysis due to its ability to extract key features from training data.
Here, let us present the super-resolution analysis as an example for extracting insights from two- and three-dimensional fluid flow data via the CNN.

\subsubsection{Two-dimensional flows}

We introduce the concept of super-resolution analysis using two-dimensional flows following Fukami {\it et al.} \cite{FFT2019}.  
The same data set of cylinder wake from the previous example is considered as a laminar flow example.  
We further consider the two-dimensional decaying homogeneous isotropic turbulence to examine the validity of the method for chaotic turbulent flows.  

The turbulent data sets are obtained from direct numerical simulation over a bi-periodic domain \cite{TNB2016}.  
The governing equation is the two-dimensional vorticity transport equation 
\begin{equation}
\frac{\partial \omega}{\partial t}+{\bm u}\cdot\nabla\omega=\frac{1}{Re}\nabla^2\omega,
\end{equation} 
where ${\bm u}=(u,v)$ and $\omega$ are the velocity and vorticity variables, respectively.
The size of computational domain and the number of grid points are $(L_x, L_y) = (1, 1)$ and $(N_x, N_y) = (128, 128)$, respectively.  
The Reynolds number $Re \equiv u^*l^*/\nu$, where $u^*$ is the characteristic velocity based on the square root of the spatially averaged initial kinetic energy, $l^*$ is the initial integral length, and $\nu$ is the kinematic viscosity.  
The initial Reynolds number $Re(t_0)=74.6$ and $\Delta t=1.950\times 10^{-4}$ are used for the present study.  
{ Here, we prepare 10000 snapshots, 70\% of which are utilized for training and remaining 30 \% are utilized for validation.}

\begin{figure}
	\vspace{0mm}
	\begin{center}
		\includegraphics[width=0.70\textwidth]{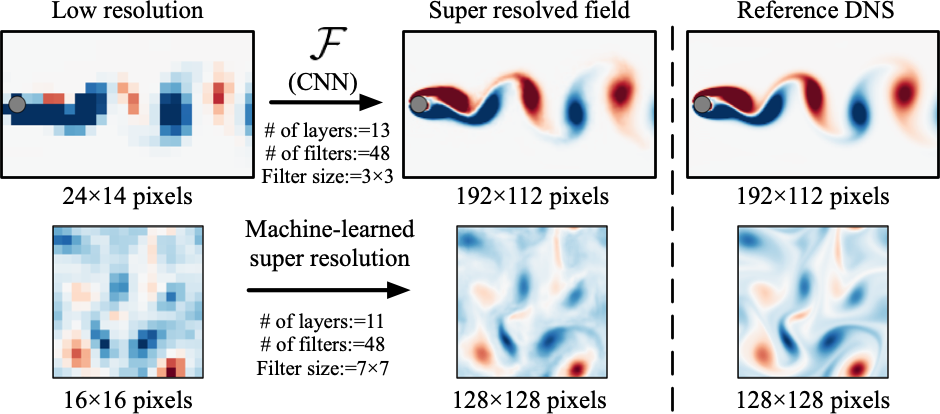}
		\caption{Super-resolution reconstruction with machine learning.  The vorticity field $\omega$ are used as input and output attributes.}
		\label{fig13}
	\end{center}
\end{figure}

For the purpose of this study, we prepare, for the input coarse images, low-resolution data by max or average pooling operations which are common in image processing.  
These operations with two-dimensional image are derived as 
\begin{equation}
 	q_{ij}^\text{LR} = \bigg[ \frac{1}{M^2} \sum_{p,s \in {\mathcal B}_{i,j}} \left( q_{ps}^\text{HR} \right)^{\mathcal B} \bigg]^{\frac{1}{\mathcal B}},
\end{equation}
where ${\mathcal B} = 1$ and $\infty$ indicate average and max pooling, respectively.  
Max pooling is able to enhance the color effects and brightness.  
With average pooling, we can extract the average value from an arbitrary area.  
The original image of $P \times Q$ pixels can be reduced to $(P/R) \times (Q/S)$ pixels.  
In the two-dimensional flow examples, $R=S=8$ with max and average pooling are adopted for the cylinder wake and decaying turbulence, respectively.

The flow chart and results of machine-learned super-resolution reconstruction are illustrated in figure \ref{fig13}.  
{ For the two-dimensional turbulence case, the test data is prepared with the initial Reynolds number of $Re(t_0)=87.7$, excluding the training process.
The details can be found in \cite{FFT2019}.}
The aforementioned coarse input data are fed into a machine learning model.  
In this example, we use a convolutional neural network for $\cal F$.  
The optimized parameters are also shown in this schematic.  
In summary, the machine-learned super-resolution problem can be written as ${\bm q}_{\rm HR} = {\cal F}({\bm q}_{\rm LR})$, where ${\bm q}$ is the feature vector.  
As the ${\bm q}$ of the two-dimensional setup, we use the vorticity field $\omega$.

As shown in figure \ref{fig13}, the reconstructed flow fields show excellent agreement with the reference DNS data in both laminar and turbulence flow examples.  
Note that if the input data is further coarse such that $R=S>8$, we need to modify the machine learning model to incorporate the physical characteristics.

\begin{figure}
	\vspace{0mm}
	\begin{center}
		\includegraphics[width=0.80\textwidth]{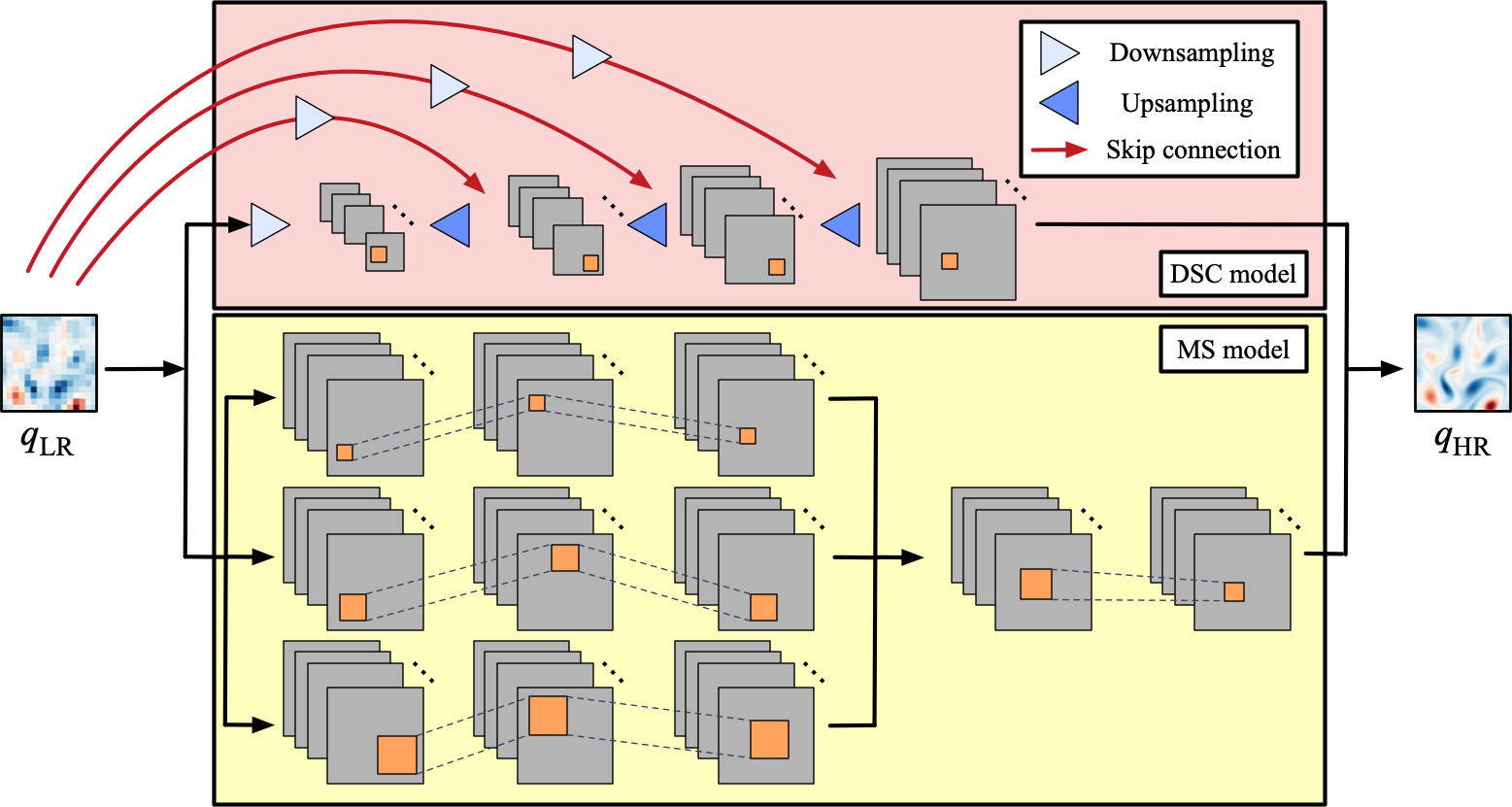}
		\caption{The two-dimensional hybrid downsampled skip-connection/multi-scale (DSC/MS) model.}
		\label{fig17_20190930}
	\end{center}
\end{figure}

Here, let us present the modified CNN called the hybrid downsampled skip-connection/multi-scale (DSC/MS) model developed by Fukami {\it et al.} \cite{FFT2019} in figure \ref{fig17_20190930}.  
As shown, the model is constructed from two parts: namely, the downsampled skip-connection (DSC) model illustrated in red, and the multi-scale (MS) model illustrated in yellow.  
The DSC model retains a robustness against rotation and translation of flow images by incorporating downsampling and skip connections into its architecture.  
On the other hand, the MS model inspired by Du {\it et al.} \cite{DQHG2018} is utilized to take the length scale of flows into account for the flow structures through a number of different filter sizes.  
We refer readers to Fukami {\it et al.} \cite{FFT2019} for details.

\begin{figure}
	\vspace{0mm}
	\begin{center}
		\includegraphics[width=0.60\textwidth]{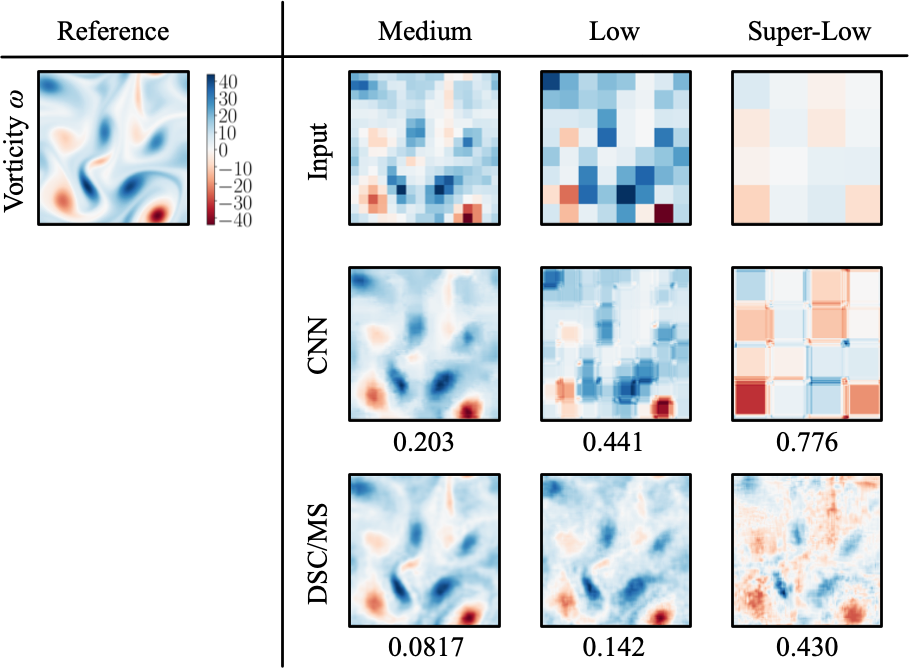}
		\caption{Super-resolution reconstruction with machine learning of two-dimensional turbulence.  Reprinted with permission from Cambridge University Press.}
		\label{fig18_20190930}
	\end{center}
\end{figure}

As the example of the super-resolution analysis, we consider cases with $R=S>8$ for the aforementioned machine learning models. 
We summarize the vorticity fields of two-dimensional decaying turbulence with $L_2$ error norm listed below the contours in figure \ref{fig18_20190930} \cite{FFT2019}.  
The flow fields reconstructed by the machine-learned models are in great agreement with the reference DNS data.  
Note that the conventional CNN does not do a good job.
It is striking that the two-dimensional turbulent flow field of $128\times 128$ pixels can be reconstructed from as little as $4\times 4$ pixels by using the proposed model.
Moreover, the machine-learned super-resolution analysis is able to increase the maximum wavenumber of the resolved flow field by approximately five fold.

\subsubsection{Three-dimensional turbulent channel flow}

As the demonstration to handle large three-dimensional data with the CNN, we consider the three-dimensional turbulent channel flow.  
The data sets are obtained by direct numerical simulation \cite{FKK2006}, for which the governing equations are the incompressible Navier--Stokes equations,
\begin{eqnarray} 
\bm{\nabla} \cdot {\bm u} = 0\\
{ \frac{\partial {\bm u}}{\partial t}  + \bm{\nabla} \cdot ({\bm u \bm u}) =  -\bm{\nabla} p  + \frac{1}{{Re}_\tau}\nabla^2 {\bm u}},
\end{eqnarray}
where $\displaystyle{{\bm u} = [u~v~w]^{\mathrm T}}$ represents the velocity with $u$, $v$ and $w$ being the streamwise ($x$), wall-normal ($y$) and spanwise ($z$) components. 
Here, $p$ is pressure, $t$ is time, and $\displaystyle{{Re}_\tau = u_\tau  \delta/\nu}$ is the friction Reynolds number.  
The quantities are non-dimensionalized using the channel half-width $\delta$ and the friction velocity $u_\tau$.  
The size of the computational domain and the number of grid points are $(L_{x}, L_{y}, L_{z}) = (4\pi\delta, 2\delta, 2\pi\delta)$ and $(N_{x}, N_{y}, N_{z}) = (256, 96, 256)$, respectively.  
The grid is taken to be uniform in all directions.  
The no-slip boundary condition is imposed on the walls and the periodic boundary condition is applied in $x$ and $z$ directions.  
These computations are conduced under a constant pressure gradient at ${Re}_{\tau}=180$.  

As the training data, we take a subdomain of the whole computational domain: the extracted domain size and number of the grid points are ($L_x^*=2\pi\delta, x\in[\pi\delta,3\pi\delta]$, $L_y^*=\delta, y\in[0,\delta]$, $L_z^*=\pi\delta, z\in[0.5\pi\delta,1.5\pi\delta]$) and $(N_{x}^*, N_{y}^*, N_{z}^*) = (128, 48, 128)$.  
Since turbulence statistics should be symmetric in $y$ direction and homogeneous in the $x$ and $z$ directions, this subdomain retains the turbulent characteristics of the channel flow. 
Similar to the two-dimensional examples, we adopt the average pooling operation to obtain the coarse input data.  
The original data of $128 \times 48 \times 128$ grids are reduced to $16 \times 6 \times 16$ grids (medium resolution) and $8 \times 3 \times 8$ grids (low resolution).
For training the machine learning model, 100 snapshots are employed.  
The time step between snapshots is $\Delta t^+ = 13.5$ in viscous time unit.
We extend the hybrid DSC/MS model for the three-dimensional super-resolution reconstruction.  
Wider range of spatial scales are present in this three-dimensional high-Reynolds number turbulent channel flow, making the present problem more challenging than the above two-dimensional flows.  
The basic parameters are same with the two-dimensional models developed by Fukami {\it et al.} \cite{FFT2019}. 

\begin{figure}
	\vspace{0mm}
	\begin{center}
		\includegraphics[width=0.80\textwidth]{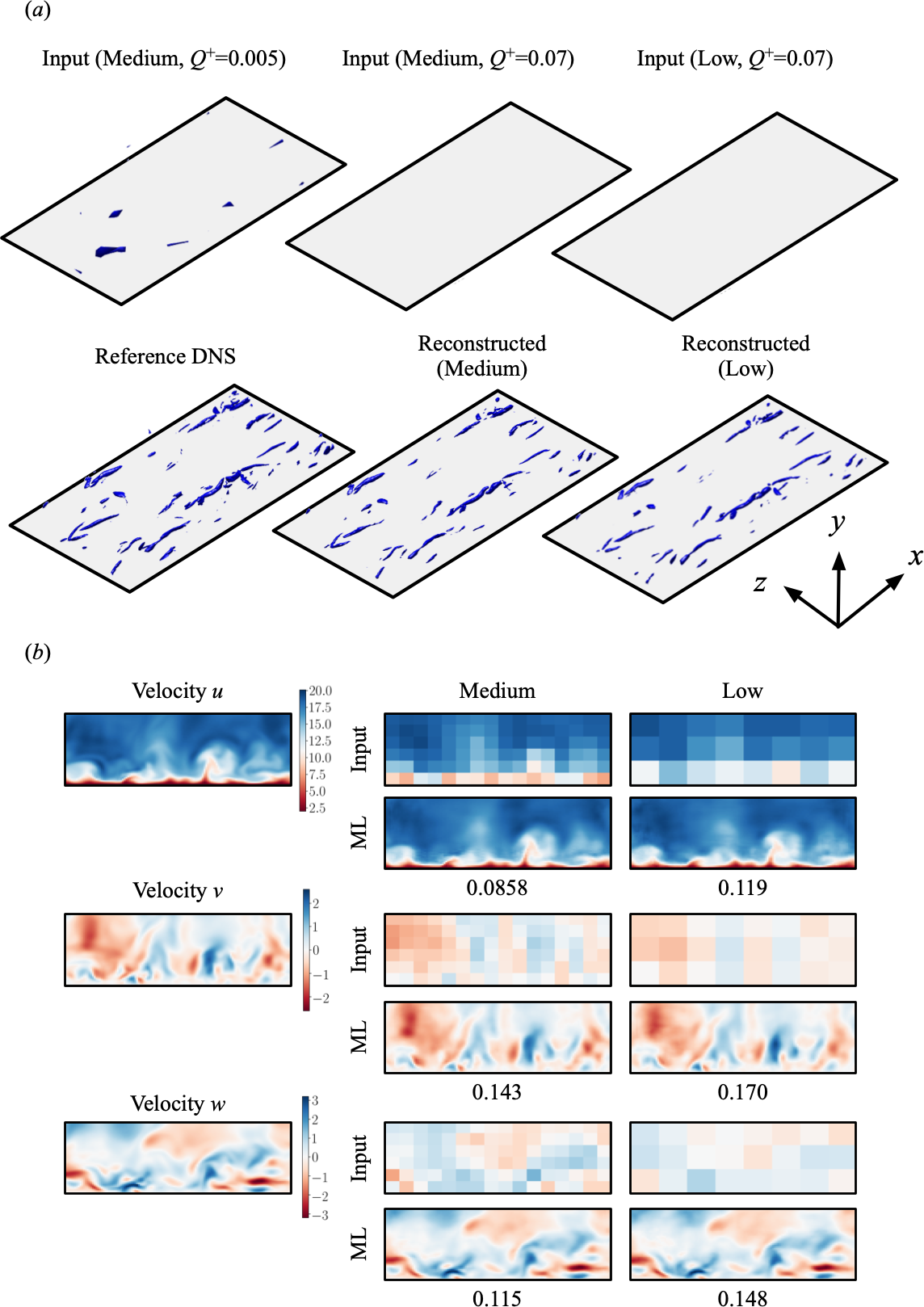}
		\caption{
		$(a)$ The $Q$-criterion ($Q^+=0.07$) contour plots of $(a)$ input coarse data, $(b)$ reference DNS data, and machine-learned super-resolved data from $(c)$ medium- and $(d)$ low-resolution inputs. $(b)$ The $y$-$z$ section contours of velocity $u$, $v$, and $w$ are for $x^+=1127$.}
		\label{fig17}
	\end{center}
\end{figure}

Let us present the $Q$-criteria isosurface contours $Q^+=0.07$ of the reference DNS data and reconstructed field in figure \ref{fig17}$(a)$.
The input coarse data contains barely any information on the vortex cores in streamwise direction.  
Almost all features are lost by the pooling operation even with medium resolution input.
{Note that the medium and low coarse input data also contain the information of vortex core.
We can find the vortex-like structures with lower threshold, e.g., $Q^+=0.005$, as shown in the upper left of figure \ref{fig17}$(a)$.}
In spite of this grossly coarse input data, the flow field can be reconstructed by the machine learning models, even for the low-resolution input.  
The velocity contours of $y-z$ section at $x^+$= 1127 are shown in figure \ref{fig17}$(b)$.  The values listed below the contour plots are the $L_2$ error norms $\epsilon_f^{\prime}=||f^{\prime}_{\rm DNS}-f^{\prime}_{\rm ML}||_2/||f^{\prime}_{\rm DNS}||_2$,
where $f^\prime$ is the fluctuation component of $f$.
The reconstructed flow fields show excellent agreement with the reference DNS data for all velocity components.  
In terms of the $L_2$ error norm, the accuracy of streamwise velocity is higher than other two components because of the strength of the feature in channel flow.

In summary, the high-resolution three-dimensional turbulent flow fields can be reconstructed from coarse flow field data through the supervised machine learning technique.  
The CNN is known to handle an input-output analysis with high-dimensional data set due to its use of filter operations.  
These results show that machine learning algorithms hold great potentials to serve as powerful analysis tools if used correctly.

\section{Concluding remarks}
\label{sec:conclude}

We assessed the performance of a number of supervised machine learning models for a range of representative regression problems for fluid flows.  
Four machine learning algorithms of multi-layer perceptron, random forest, support vector regression, and extreme learning machine were considered for laminar flow problems.  
As shown in sections 3.1 and 3.2, users can select a network from a zoo of machine-learning models \cite{BK2019}, taking into account the computational cost, accuracy, robustness, and training data density.  

Furthermore, we considered problems that require the handling of two- or three-dimensional fluid flow data.  
In particular, we considered the super-resolution analysis with convolutional neural network.
We found that convolutional neural network is a powerful tool to handle big data, which capitalizes on the concept of filter operation to beat the curse of dimensionality.  
An adjustable filter for non-uniform grids including body-fitted grids used in high-$Re$ airfoil flow may be desirable for future applications in fluid flows.  

The field of supervised machine learning for fluid dynamics can be combined with other {\it data-oriented} architecture including linear-based methods \cite{TBDRCMSGTU2017,THBSDBDY2019}, network science \cite{GMNT2018}, and artificial intelligence.  
Prior to employing machine learning, users should care what type of fluid flow problems can really benefit from supervised machine learning with fluid dynamics.  
Although neural networks are known as an ``universal approximator"\cite{Kreinovich1991,Hornik1991,Cybenko1989,BFK2018}, the supervised machine learning models are not suitable for extrapolation problem in terms of training data \cite{Taira2019}.  
While it is presently difficult to define the border of interpolation and extrapolation for fluid flow applications, future efforts may enable us to establish machine-learning models for fluid flows that can warn us if a model attempts to step outside of the demarcation of training data boundary.

\begin{acknowledgements}

Kai Fukami and Koji Fukagata thank the support from the Japan Society for the Promotion of Science (KAKENHI grant number: 18H03758).  Kunihiko Taira acknowledges the support from the US Army Research Office (grant number: W911NF-19-1-0032) and the US Air Force Office of Scientific Research (grant number: FA9550-16-1-0650).  The authors thank Mr.~Muralikrishnan Gopalakrishnan Meena (University of California, Los Angeles) for sharing his DNS data.

\end{acknowledgements}

\end{document}